# Nanoscale Analysis of Frozen Water by Atom Probe Tomography Using Graphene Encapsulation and Cryo-Workflows: A Parametric Study

Florant Exertier[1,*], Levi Tegg[2,3], Adam Taylor[1], Julie M. Cairney[2,3], Jing Fu[4] and Ross K.W. Marceau[1,*]

[1] Deakin University, Institute for Frontier Materials, Geelong, VIC 3216, Australia

[2] School of Aerospace, Mechanical and Mechatronic Engineering, The University of Sydney, Sydney, NSW 2006, Australia

[3] Australian Centre for Microscopy and Microanalysis, The University of Sydney, Sydney, NSW 2006, Australia

[4] Department of Mechanical and Aerospace Engineering, Monash University, Clayton, VIC 3800, Australia

*Corresponding authors: Florant Exertier, f.exertier@research.deakin.edu.au;

Ross K.W. Marceau, ross.marceau@deakin.edu.au.

Deakin University, Institute for Frontier Materials, 75 Pigdons Road, Waurn Ponds, VIC 3216, Australia. Phone: +61 3 52271283

## Abstract

There has been an increasing interest in atom probe tomography (APT) to characterise hydrated and biological materials. A major benefit of APT compared to microscopy techniques more commonly used in biology is its combination of outstanding 3D spatial resolution (~0.2 nm) and mass sensitivity. APT has already been successfully used to characterise biological materials, revealing key structural information at the atomic scale, however there are many challenges inherent to the analysis of hydrated materials. New preparation protocols, often involving sample preparation and transfer at cryogenic temperature, enable APT analysis of hydrated materials and have the potential to enable 3D atomic scale characterisation of biological materials in the near-native hydrated state. In this study, APT specimens of pure water at the tips of tungsten needles were prepared at room temperature by graphene encapsulation. A parametric study was conducted where samples were transferred at either room temperature or cryo-temperature and analysed by APT by varying parameters such as the flight path and pulsing mode. The differences between the acquisition scenarios are presented along with recommendations for future studies.

## Introduction

There is an acute need to understand complex hydrated biological systems. Near-native state imaging of most biological specimens involves maintaining the samples in a hydrated state, which often requires the use of cryogenic workflows. Several characterisation methods that involve cryogenic temperature preparation (e.g. plunge-freezing) have been employed, such as cryogenic transmission electron microscopy (cryoTEM) (Dubochet, et al., 1988; Patterson, et al., 2017), electron cryotomography (cryoET) (Gonen & Nannenga, 2021), and atom probe tomography (APT) (McCarroll, et al., 2020; Gault, et al., 2021). These methods usually involve rapid cooling of the hydrated sample by plunge-freezing in order to prevent the formation of crystalline ice that causes structural damage, and instead, maintain the sample in a vitreous state that preserves its near-native state (Dubochet, et al., 1988; Marko, et al., 2007). To lower the risk of any damage to frozen samples, they are usually stored and transferred using dedicated cryogenic transfer tools. Whilst this type of protocol has been routinely used for several decades to analyse hydrated biological materials using cryoEM (Dubochet, et al., 1988; Marko, et al., 2007; Gonen, et al., 2021; Parmenter & Nizamudeen, 2021), cryo-APT is still under development (Schreiber, et al., 2018; Chen, et al., 2021; Stender, et al., 2022). APT is the only technique that can enable three-dimensional sub-nanometre scale compositional and structural analysis (Gault, et al., 2021).

Encouraged by the interest in achieving APT analysis of near-native state hydrated biological materials, as well as the progress of cryo-EM in the field, a number of APT studies have already been published where pure water samples were successfully field evaporated (El-Zoka, et al., 2020; Schwarz, et al., 2020; Stender, et al., 2022). Although hydrated materials have been studied using field evaporation techniques as early as 1968 (Anway, 1969), APT of hydrated materials remains challenging for several reasons, the main being the stringent conditions required during an atom probe experiment (i.e., ultra-high-vacuum ($< 1\times10^{-10}$ torr) and cryogenic temperature) (Stender, et al., 2022). To date, the hydrated materials studied using APT include aqueous solutions (Stephenson, et al., 2018; Qiu, et al., 2020b; Qiu, et al., 2020c; Schwarz, et al., 2021), liquid/solid interfaces (Schreiber, et al., 2018; Perea, et al., 2020; Qiu, et al., 2020d) and pure water (El-Zoka, et al., 2020; Qiu, et al., 2020b; Schwarz, et al., 2020; Stender, et al., 2022), and they were usually analysed using laser-pulsed APT. The samples were prepared using either graphene encapsulation, reported to produce vitreous water ice (Zhang, et al., 2022), or encapsulation using nano-porous vessels and cryo-FIB milling, likely producing crystalline ice without evidence to the contrary. Other methods to obtain vitreous ice APT needles have also recently been explored (Zhang, et al., 2022).

Here, graphene-coating has been used to encapsulate a small volume of water between the tip of a tungsten (W) needle substrate and several single graphene layers, and analysed with different APT acquisition parameters. To study the influence of the method of water freezing on the field evaporation behaviour of graphene-encapsulated water, graphene-encapsulation was combined with cryo-workflows involving plunge-freezing, cryo-FIB removal of the graphene layers, and cryo-transfer between instruments. In addition to studying the structure and chemistry of graphene-encapsulated water and understanding the field evaporation of small volumes of water, this study provides recommendations for future work involving the study of liquid samples (e.g., solutions, suspensions) or hydrated samples.

## Materials and Methods

Samples made by graphene-encapsulation of water at the tip of W needles have been analysed using APT under different analysis conditions summarised in **Table 1**, by varying sample preparation temperature, instrument flight path (reflectron-fitted LEAP, or straight flight path LEAP) and pulsing mode (voltage or laser). The employed workflows are illustrated in **Figure 1**. Regardless of the conditions, all samples were prepared following the same preliminary steps. Firstly, W support needles (ProSciTech, 99.99% purity) were prepared by electropolishing using a 5% NaOH aqueous solution. The samples were then 'pre-run', i.e., field-evaporated in the atom probe to remove impurities and clean the surface as described in (Exertier, et al., 2022). Pure (Milli-Q) water was then trapped at the tip of the sharp needles using graphene-encapsulation, as described in (Qiu, et al., 2020b; 2020c; 2020d; Exertier, et al., 2022). Finally, SEM imaging was performed to check the surface of the graphene coating layers. For the room temperature (RT) experiments the samples were transferred directly to the atom probe for analysis, but the cryo-temperature (CT) samples underwent several additional steps, as described in **Figure 1**, using cutting-edge equipment and novel laboratory protocols, which are presented in the supplementary materials (Section 3: Cryo workflow).

**Table 1** Summary of the different employed experimental conditions, referring also to **Figure 1**. The number of data sets collected for each scenario are noted in parentheses. The tick marks indicate whether a scenario/pulsing mode combination was attempted or not.

| Instrument | LEAP 5000 XR | LEAP 4000 X Si | | |
|---|---|---|---|---|
| Flight path (FP) type | Reflection-fitted FP | Straight FP | | |
| Sample preparation temperature (RT/CT) | Room temperature | Room temperature | Cryogenic temperature | |
| | | | No FIB (A) | FIB (B) |
| Laser-pulsed APT (no. runs) | ✓ (6) | ✓ (1) | ✓ (1) | ✓ (2) |
| Voltage-pulsed APT (no. runs) | ✓ (5) | ✓ (1) | ✓ (2) | |

Reflectron-fitted APT experiments were conducted at Deakin University's Advanced Characterisation Facility using a LEAP 5000 XR (CAMECA Instruments), and the straight flight path APT experiments were conducted on a LEAP 4000 X Si (CAMECA Instruments) located at the University of Sydney, at the Australian Centre for Microscopy and Microanalysis. The latter is equipped with a vacuum cryo-transfer interface enabling the study of cryogenically prepared samples (Cairney, et al., 2019; Chen, et al., 2020; McCarroll, et al., 2020; Chen, et al., 2021). The cryo-APT experiments were conducted in Sydney using the workflow summarised in **Figure 1**, split into two scenarios. In scenario A (referred to as cryo-A), graphene-encapsulated needles were inserted into a glovebox, then plunge-frozen using liquid nitrogen ($LN_2$). Using a vacuum cryo transfer (VCT) suitcase (Ferrovac, GmbH), the samples were then transferred to the atom probe instrument for analysis. In scenario B (referred to as cryo-B), the samples were plunge-frozen in the same fashion. However, prior to loading into the atom probe, the samples were cryo-transferred to a cryo-FIB instrument to undergo cryo-annular milling, in order to remove the graphene layer so that only pure water remained at the tip of the specimen needle.

More detailed information about the cryo-workflows and the equipment involved is presented in the supplementary materials (Section 3: Cryo workflow).

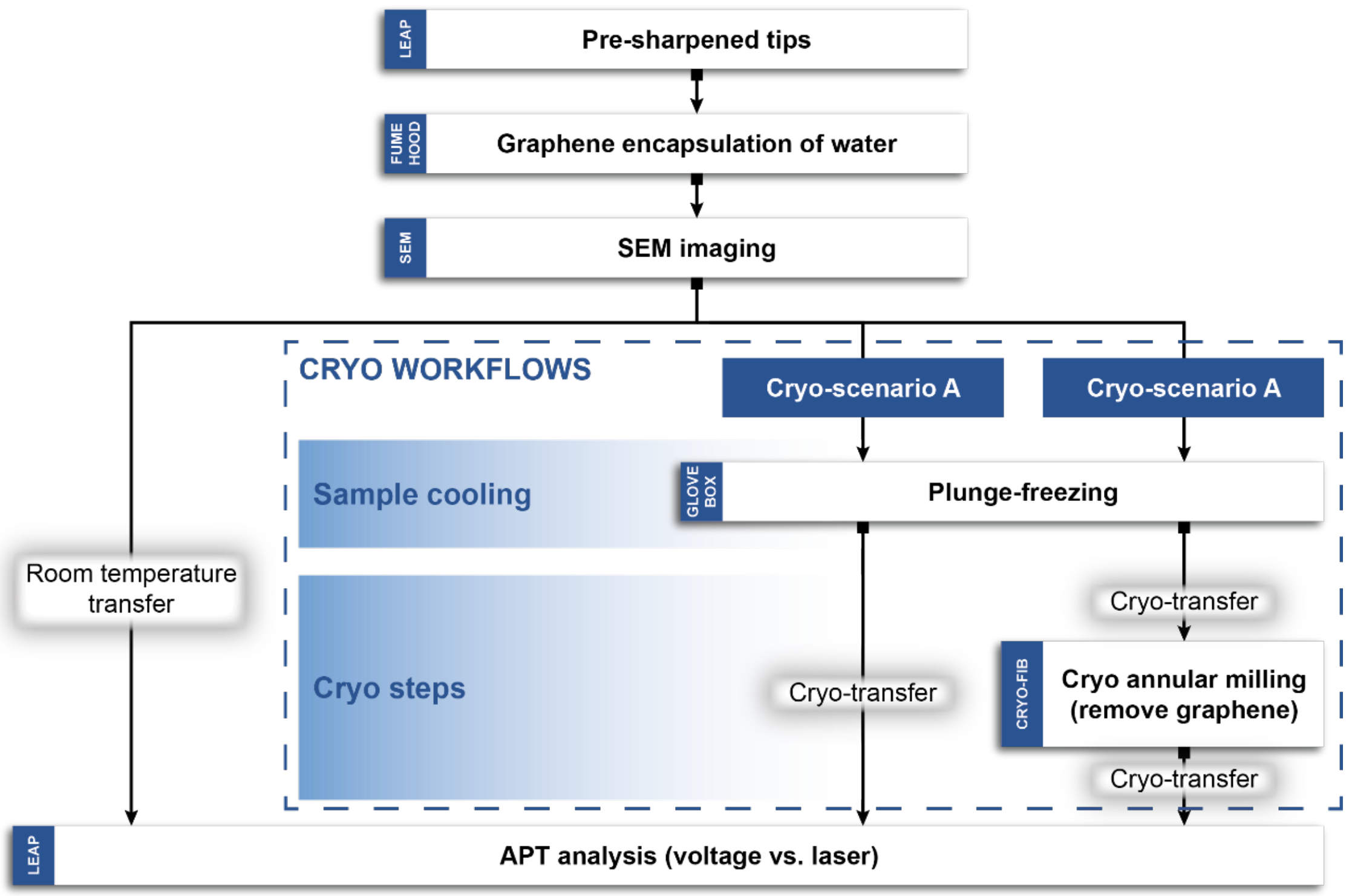

**Figure 1** Summary of the cryo-workflows.

## Results

Pre-run W needles were graphene-coated multiple times using a speed-controlled device, with a coating speed of 0.02 mm/s, then analysed using APT with different conditions summarised in **Table 1**. As mentioned in the Materials and Methods section, the tip shape of RT samples could be assessed using SEM imaging prior to APT analysis. For CT samples, whilst the shape of the cryo-B samples could be assessed during the FIB milling process, the cryo-A samples were transferred directly to the atom probe and therefore could not be imaged. In this case, the tip condition was assessed using in-situ images from the atom probe, as illustrated in **Figure**

**2**, showing three consecutive screen captures from the acquisition software taken at different times during the acquisition. Post-graphene coating SEM images are presented in the Supplementary materials and show a clean surface at the apex of the specimens. For the RT samples, these images are representative of the conditions of the samples at the start of the experiment, however for the CT samples, frost was occasionally observed, either engulfing the entire specimen tip (**Figure S16**), or built up in the form of small chunks scattered along the specimen shank, as illustrated in **Figure 2a**.

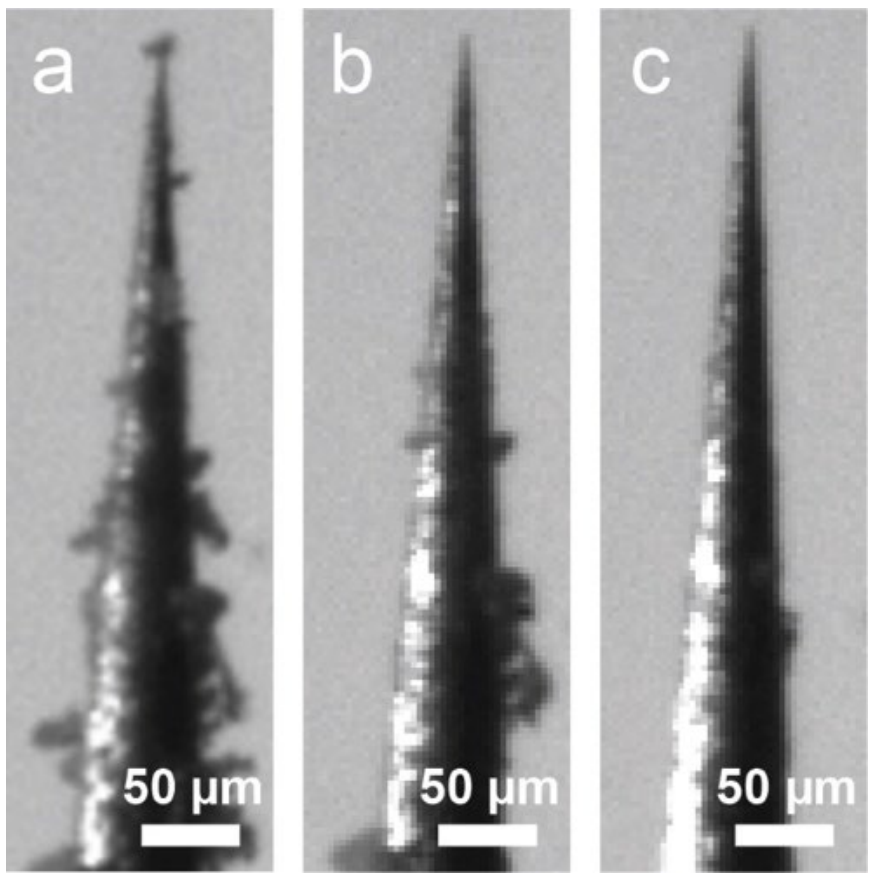

**Figure 2** In-situ optical microscope images of an APT needle prepared following the cryo-A scenario, captured (a) before starting the acquisition, (b) after 36,000 detected ions, and (c) after 123,000 detected ions.

All experiments conducted on the reflectron-fitted instrument were carried-out using a pulse frequency of 200 kHz, however, the experiments conducted on the straight FP instrument were carried out at 125 kHz to minimise the "wrap around" effect of heavier ions, as described by Larson, et al. (2013). All the datasets presented here were acquired at temperature set-point of 50 K. Some samples were analysed at 70 K to study the influence of the temperature on the experimental yield, but no improvement was reported. For all the experiments conducted on the straight FP instrument, the voltage was increased manually until mostly W ions were detected, at which stage the automatic voltage control functionality was activated (Larson, et al., 2013).

APT results from each scenario detailed in **Table 1** are presented in **Figure S1** in the supplementary materials. These results show the 3D reconstruction, detection rate and voltage history, as well as mass-to-charge ratio against the ion sequence, for each sample. The 3D reconstructions were all obtained using the “Tip Profile” method, using pre-APT SEM images of the graphene-coated needles. The data from (1) the graphene coated specimen prepared at room temperature and analysed using the reflectron-fitted atom probe using laser pulses, and from (2) the graphene coated specimen prepared using cryo-scenario A and analysed using the straight flight path instrument using voltage pulses are presented in **Figure 3** and **Figure 4**, respectively.

As shown in **Figure 3a** and **Figure 4a**, the region of interest (ROI) was located at the tip of the specimen needle, however the experiments were interrupted after the total detected ion count was deemed large enough for data analysis (typically > 500,000 ions). While the 3D reconstructions represent the entire datasets, the detection rate/voltage plots (**Figure 3b** and **Figure 4b**) only show the first $3\times10^5$ ions, and the mass history plots (**Figure 3c** and **Figure 4c**) were further cropped to show the most relevant part of the graph represented by green dashed rectangles.

| Reflectron fitted instrument |
| --- |
| RT specimen preparation |
| **Laser pulsing** |

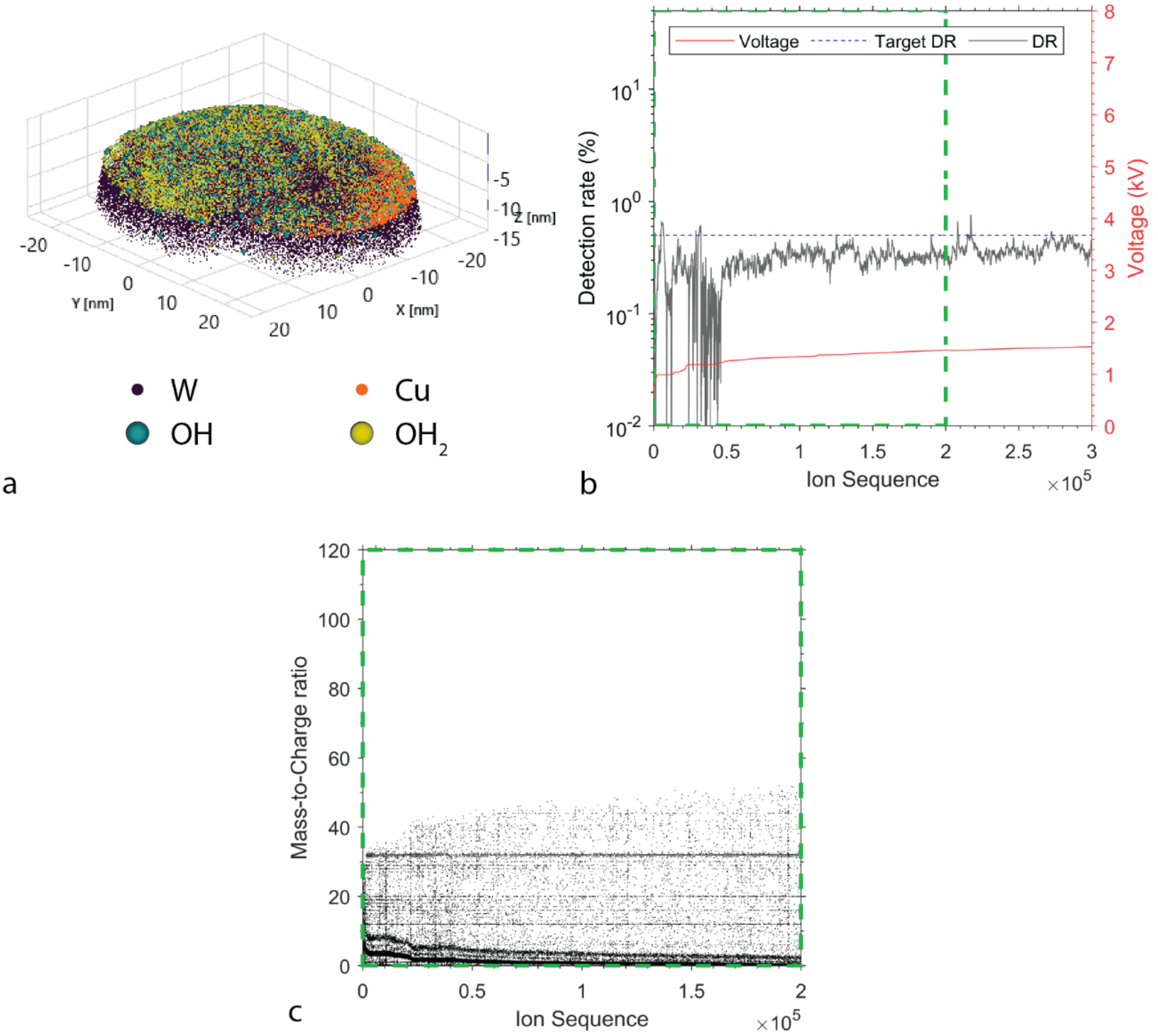

**Figure 3.** Graphene-coated specimen prepared at room temperature and analysed using the reflectron-fitted instrument in laser-pulsing mode. (a) 3D reconstruction. (b) Detection rate (DR) and voltage evolution plotted against the ion sequence. Green rectangle corresponds to the cropped window shown in (c). (c) Mass-to-charge ratio plotted against the ion sequence.

**Figure 4** shows a clear boundary between the ROI and the W substrate. The mass history plots show similar field evaporation sequences to previously published results (Qiu, et al., 2020b; Exertier, et al., 2022): the first ion types detected are water-related ions and carbon-related ions (from the graphene), next are ions corresponding to the W oxide layer, before reaching a stable regime where only pure W is detected in the 3+ and 4+ charge states, regardless of the scenario.

These three distinct stages can be seen on both the mass history plots and the voltage history plots. The region corresponding to the water-related ions usually contains a low-voltage plateau, followed by an increase in voltage during which the W oxides are detected, and finally the voltage stabilises to a final, but slightly increasing value, corresponding to the pure W material.

| Straight flight path |
| --- |
| Cryo specimen preparation - Scenario A Cryo-A (No FIB) |
| **Voltage pulsing** |

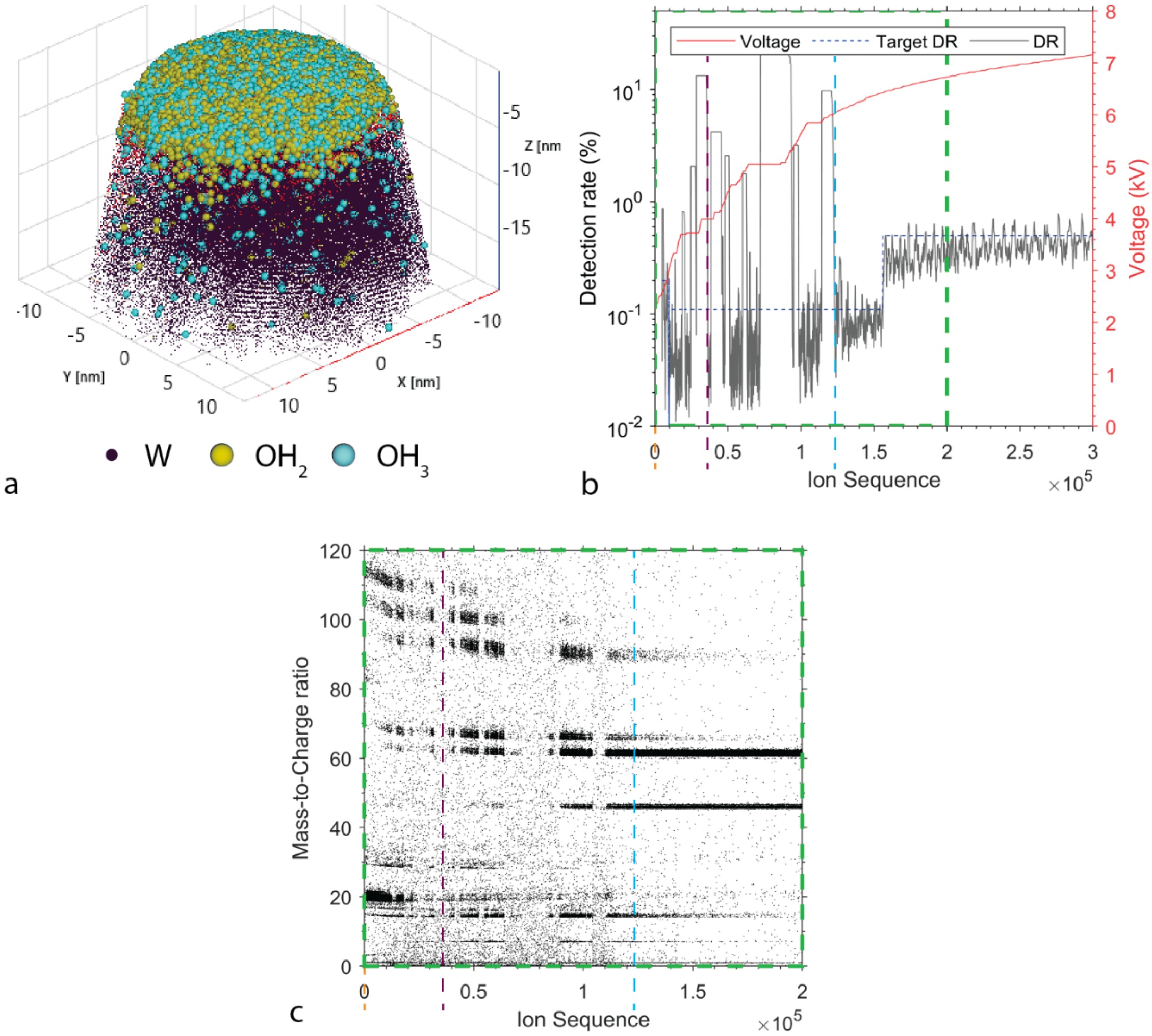

**Figure 4.** Graphene-coated specimen prepared using cryo-scenario A and analysed using the straight flight path instrument in voltage-pulsing mode. (a) 3D reconstruction. (b) Detection rate (DR) and voltage evolution plotted against the ion sequence. Green rectangle corresponds to the cropped window shown in (c). (c) Mass-to-charge ratio plotted against the ion sequence.

Mass spectra were generated for one dataset from each acquisition scenario and are shown in **Figure 5**. Mass spectrum D (straight FP, RT, VP) contains a ROI of ~4,000 ions, which is too small to be accurately compared to the other datasets (> 10,000 ions) because the signal-to-background noise ratio measured for the peaks corresponding to this ROI is too low, causing all the detected peaks to have a net count below the critical level as defined by Currie (1968) and mentioned in (Larson, et al., 2013; Exertier, et al., 2022). During the reconstruction process, the W peaks were used for the time-of-flight correction and mass calibration (Larson, et al., 2013), which sometimes resulted in a slight shift in the position of some of the smaller peaks (e.g. water-related peaks) by up to 0.3 Da. All mass spectra show W in the 3+ and 4+ charge states, as well as $^{16}O^{+}$ (16 Da), $^{16}OH^{+}$ (17 Da) and $^{16}OH_2^{+}$ (18 Da). The peaks deemed to be above the background level in the mass spectra are presented in **Table 2**. Several peaks, such as those located at 28 Da and 29 Da (identified as $CO^{+}$ and $COH^{+}$, respectively), are present in most mass spectra. The presence of Cu in the first two mass spectra (reflectron-fitted FP, LP and VP) at 31.5, 32.5, 63 and 65 Da, is due to impurities remaining in the encapsulated solution from the Cu etching process, prior to graphene-coating, as explained by Exertier, et al. (2022). While the detection of Cu in the reflectron-fitted FP datasets and the absence of detection for all straight FP datasets could be correlated to the instrument FP, it is more likely related to experimental issues that were faced during the graphene-coating etching process prior to conducting the reflectron-fitted FP experiments. These issues were related to complete removal of Cu (as-received substrate for the single layer graphene) after the solution-based graphene etching process, and were solved by the time the straight FP experiments were conducted. While the detection of Cu originating from the etching process prior to graphene-coating may suggest the contribution of N-containing ions (from the ammonium persulfate (APS) etchant solution) to the peaks at 14, 15, 28 and 29 Da, the absence of detection of Cu ions in the straight FP data suggests that in these cases the APS solution has been successfully removed, supporting their assignment as $CO^{2+}$, $COH_2^{2+}$, $CO^+$ and $COH^+$, respectively.

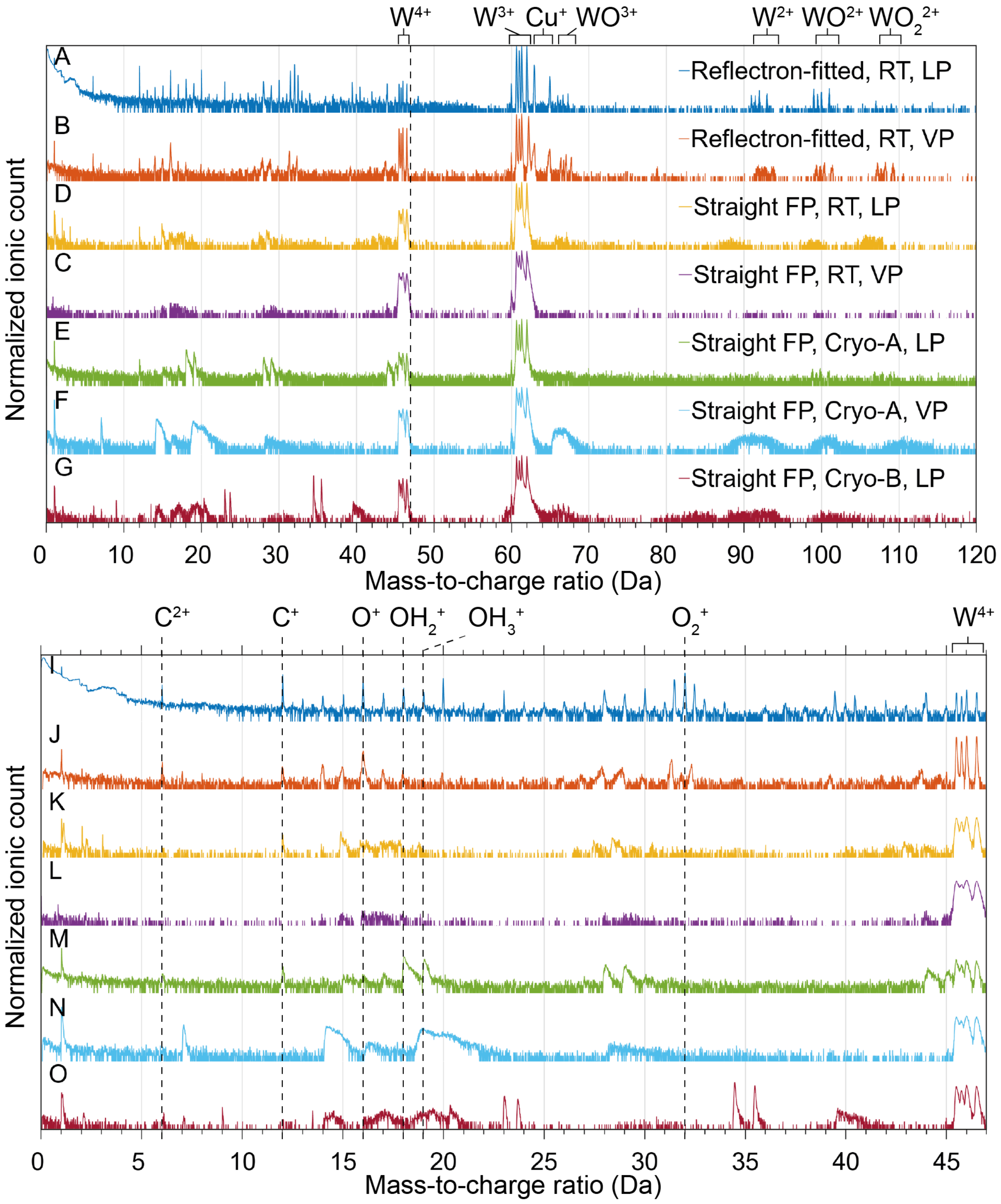

**Figure 5** Mass spectra related to each acquisition scenario as summarised in **Table 1**. Each mass spectrum was normalised to the maximum count from the tallest peak ($^{184}W^{3+}$). RT, Cryo-A and Cryo-B refer to the specimen preparation temperature as indicated in **Table 1**. (A to G) show the full mass spectra from 0 Da to 120 Da, and (I to J) show the spectra close up from 0 Da to 47 Da.

**Table 2** List of peaks for each field evaporation scenario (A to G from **Figure 5**).

| m/z (Da) | Peak assignment | A | B | C | D | E | F | G |
|---|---|---|---|---|---|---|---|---|
| 1 | $H^+$ | ✓ | ✓ | ✓ | ✓ | ✓ | ✓ | ✓ |
| 2 | $H_2^+$ | | ✓ | ✓ | | | | |
| 3 | $H_3^+$ | | | ✓ | | | | |
| 6 | $C^{2+}$ | ✓ | ✓ | | | ✓ | | ✓ |
| 7 | $N^{2+}$, $CH_2^{2+}$ | | ✓ | | | | ✓ | ✓ |
| 9 | $OH_2^{2+}$ | | | | | | | ✓ |
| 12 | $^{12}C^+$, $^{12}C_2^{2+}$ | ✓ | ✓ | ✓ | ✓ | ✓ | | |
| 12.5 | $^{12}C_2H^{2+}$, $C_2^{2+}$ $(^{12}C^{13}C)^{2+}$ | ✓ | | | | | | |
| 13 | $^{12}CH^+$, $^{13}C^+$, $CN^{2+}$, $^{13}C_2^{2+}$ | ✓ | | | | | | |
| 14 | $CO^{2+}$, $CH_2^+$, $N^+$ | ✓ | ✓ | ✓ | ✓ | | ✓ | |
| 15 | $CH_3^+$, $NH^+$ | ✓ | ✓ | ✓ | ✓ | ✓ | | ✓ |
| 16 | $O^+$ | ✓ | ✓ | ✓ | ✓ | ✓ | ✓ | ✓ |
| 17 | $OH^+$ | ✓ | ✓ | ✓ | ✓ | ✓ | ✓ | ✓ |
| 18 | $OH_2^+$ | ✓ | ✓ | ✓ | ✓ | ✓ | ✓ | ✓ |
| 19 | $OH_3^+$ | ✓ | | ✓ | ✓ | ✓ | ✓ | ✓ |
| 20 | $C_2O^{2+}$ | ✓ | ✓ | | | | | ✓ |
| 23 | $^{23}Na^+$, $^{69}Ga^{3+}$ | ✓ | | | | | | ✓ |
| 23.67 | $^{71}Ga^{3+}$ | | | | | | | ✓ |
| 24 | $^{12}C_2^+$ | ✓ | | | | | | |
| 25 | $^{12}C_2H^+$, $C_2^+$ $(^{12}C^{13}C)^+$ | ✓ | | | | | | |
| 26 | $C_2H_2^+$, $CN^+$, $^{13}C_2^+$ | ✓ | | | | | | |
| 27 | $C_2H_3^+$, $CNH^+$ | ✓ | ✓ | | | | | |
| 28 | $CO^+$, $C_2H_4^+$, $N_2^+$ | ✓ | ✓ | ✓ | ✓ | ✓ | ✓ | |
| 29 | $COH^+$, $C_2H_5^+$ | ✓ | ✓ | ✓ | ✓ | ✓ | ✓ | |
| 30 | $COH_2^+$, $C_2H_6^+$ | ✓ | ✓ | | | | | |
| 31 | $COH_3^+$, $C_2H_7^+$ | ✓ | | | | | | |
| 31.5 | $^{63}Cu^{2+}$ | ✓ | ✓ | | | | | |
| 32 | $O_2^+$ | ✓ | ✓ | | | | | |
| 32.5 | $^{65}Cu^{2+}$ | ✓ | ✓ | | | | | |
| 33 | $O_2H^+$ | ✓ | | | | | | |
| 34 | $O_2H_2^+$ | ✓ | | | | | | |
| 34.5 | $^{69}Ga^{2+}$ | | | | | | | ✓ |
| 35.5 | $^{71}Ga^{2+}$ | | | | | | | ✓ |
| 39.5 | $^{63}CuO^{2+}$ | ✓ | | | | | | |
| 40.5 | $^{65}CuO^{2+}$ | ✓ | | | | | | |
| 44 | $CO_2^+$ | ✓ | ✓ | ✓ | ✓ | ✓ | | |
| 45 | $^{180}W^{4+}$, $CO_2H^+$ | ✓ | ✓ | ✓ | ✓ | ✓ | ✓ | ✓ |
| 45.5 to 46.5 | $^{182\text{-}186}W^{4+}$ | ✓ | ✓ | ✓ | ✓ | ✓ | ✓ | ✓ |
| 60 to 62 | $^{180\text{-}186}W^{3+}$ | ✓ | ✓ | ✓ | ✓ | ✓ | ✓ | ✓ |
| 63 | $^{63}Cu^{+}$ | ✓ | ✓ | | | | | |
| 65 | $^{65}Cu^{+}$ | ✓ | ✓ | | | | | |
| 66 to 67.3 | $^{182\text{-}186}WO^{3+}$ | ✓ | ✓ | ✓ | | | ✓ | |
| 91 to 93 | $^{182\text{-}186}W^{2+}$ | ✓ | ✓ | | | | ✓ | |
| 99 to 101 | $^{182\text{-}186}WO^{2+}$ | ✓ | ✓ | ✓ | | ✓ | ✓ | |
| 107 to 109 | $^{182\text{-}186}WO_2^{2+}$ | ✓ | ✓ | ✓ | | ✓ | ✓ | |

As mentioned previously, the field evaporation sequences displayed in **Figure S1** show three different evaporation stages, the first two corresponding to the water phase and to the W oxide phase, respectively. The fact that the W oxide ions only appear after the standing voltage reaches a certain value could be related to their high mass-to-charge (m/z) ratio being out of the m/z window, which is determined by the standing voltage and pulse frequency, as explained by Larson, et al. (2013). At a lower voltage, field-evaporated ions have a longer time-of-flight (TOF), which decreases as the voltage increases. At the earliest stage of field evaporation (low voltage), some heavy ions may have a TOF longer than the duration between pulses, resulting in these ions not being detected within the window of time for detection corresponding to that pulse. In some cases, a “wrap-around” effect may be observed in the mass spectrum as a broad peak at a lower m/z range, caused by these heavy ions being detected during the time window corresponding to the next or later pulse. In this work, the absence of W oxide ions at the earliest stage of the field evaporation could be explained by this phenomenon, especially in the case of reflectron-fitted FP experiments, where the voltage at which the water-related ions field evaporate is below 2 kV. In (Larson, et al., 2013), the maximum m/z range at 2 kV with a 382 mm FP instrument and a 200 kHz pulse frequency is estimated to be ~88 Da, which is lower than that of the W oxide ions in the 2+ charge state (≥ 99 Da for $^{182}WO^{2+}$). However, for similar voltages and pulse frequencies, the maximum m/z range is larger when the FP is shorter. In the same conditions (2 kV, 200 kHz) a 90 mm FP instrument has a maximum m/z range of ~2015 Da (Larson, et al., 2013). Despite this much greater m/z range for the straight FP instrument used in this study (90 mm FP), the same field evaporation stages were observed, where the water-related ions were detected first, followed by the W oxide phase. This suggests that these apparent stages depict the accurate position of these phases along the z-depth of the APT needles. The maximum m/z ranges at a given pulse frequency are shown in **Figure 6** where the full mass history and voltage history are plotted as a function of the evaporation sequence for experiments conducted using both pulsing modes (laser and voltage) on both FP type instruments. The maximum observed m/z range are also represented as a function of the standing voltage and compared to the estimations provided by Larson, et al. (2013) in **Figure 6e**. It should be noted that in **Figure 6d** the sudden drop in m/n maximum range at 120,000 ions is due to a manual increase of the pulse frequency by the operator from 125 kHz to 200 kHz to accelerate the field evaporation after the pure W phase was reached. Incidentally, this illustrates the reduced m/z detection window for a given voltage when the pulse frequency is increased. Similarly, the changes in apparent noise density in **Figure 6b** and **Figure 6c** at 50,000 ions and 100,000 ions, respectively, were caused by a manual increase of the detection rate by the operator for the same reason.

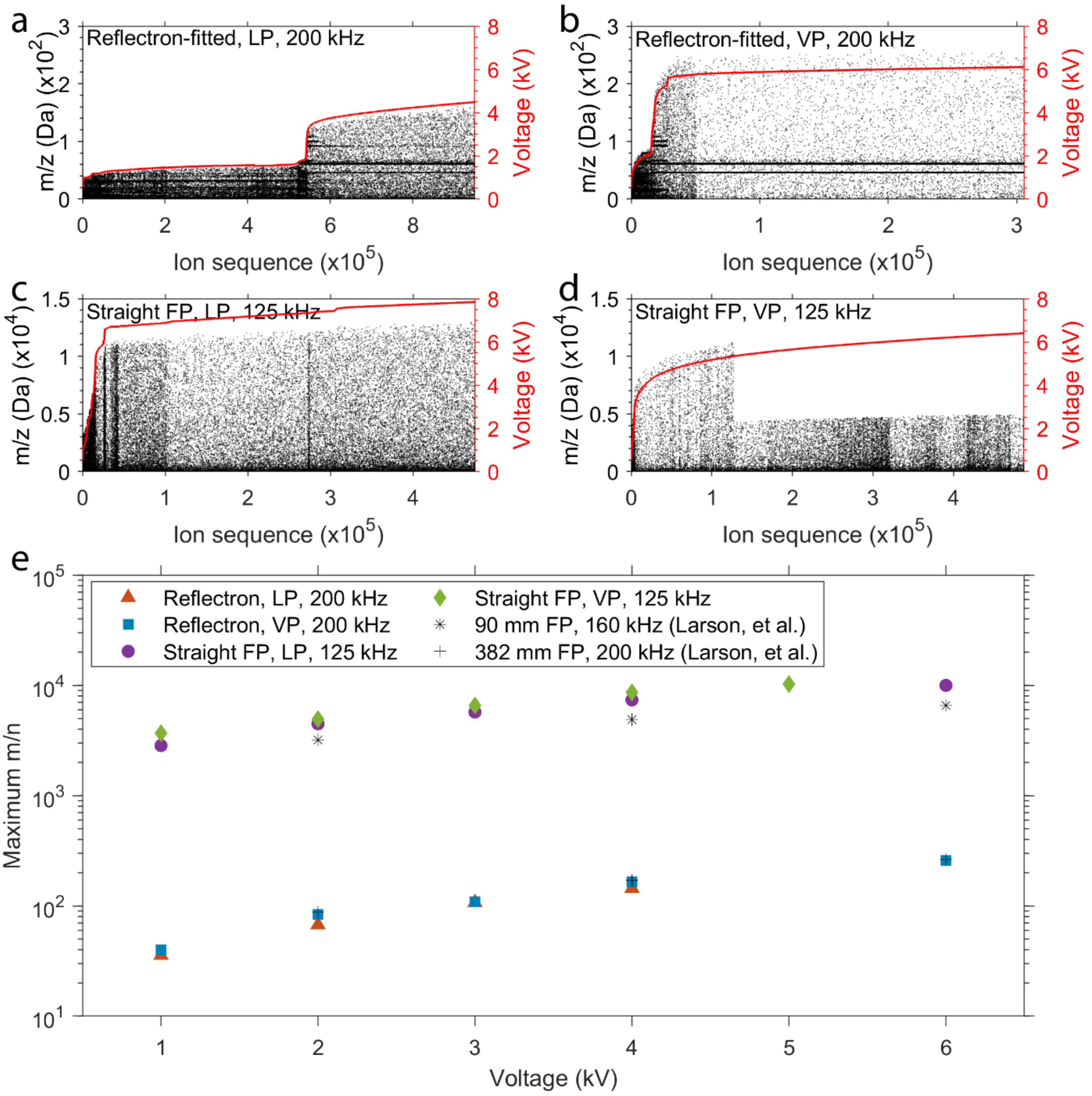

**Figure 6** (a-d) Standing voltage and mass-to-charge ratio history as a function of the ion sequence for experiments conducted on the reflectron-fitted instrument using (a) laser pulsing and (b) voltage pulsing, and on the straight FP instrument using (c) laser pulsing, and (d) voltage pulsing. (e) Maximum observed m/z value for each experiment at given voltages compared to similar observation from Larson, et al. (2013) at 200 kHz.

Several data quality metrics were measured and compared in **Table 3**. The mass resolving power (MRP) was measured by calculating the ratio of the counts at the full width at half maximum to the peak position (Gault, et al., 2012) for the $^{184}W^{3+}$ peak (61.3 Da) and the $^{16}OH_2^+$ peak (18 Da). For the former peak, the MRP was consistently higher on the reflectron-fitted instrument than on the straight FP instrument, as expected. For each scenario where both voltage and laser pulsing were tested, the MRP for the laser-pulsed experiment was always higher than that of the voltage-pulsed experiment. Similarly, the signal-to-background (STB) was higher by one order of magnitude for reflectron-fitted FP data than for straight FP data, although no significant difference was discerned between laser and voltage pulsing. The same observations were made for the $^{16}OH_2^+$ peak, except for the data acquired on the straight FP instrument for samples prepared at room temperature, for which the amount of water detected was too small to be accurately compared. **Table 3** also shows the $H_3O^+$ to $H^+$ concentration ratio and an estimate of the electrostatic field. The latter was calculated using the Kingham curves for tungsten (Kingham, 1982), as shown in **Figure S2**. It should be noted that this estimate is mostly representative of the field conditions near the apex of the specimen when the surface ions (i.e. graphene and water-related ions) have been evaporated and only W is left. Prior to reaching the W substrate, the electrostatic field is likely lower, as reported in our previous study (Exertier, et al., 2022), and increasing as suggested by the voltage curve profile of each experiment (**Figure S1**).

**Table 3** Mass resolving power (MRP), signal-to-background (STB) ratio and charge-state ratio (CSR) calculated for two peaks for each experimental scenario. A field estimate calculated using the $W^{4+}/(W^{3+}+W^{4+})$ CSR is also shown, as well as the $H_3O^+$ to $H^+$ concentration ratio.

| Flight path | Prep. temp. | Pulse mode | Ion count ×10$^6$ | $^{184}W^{3+}$ MRP ×10$^2$ | $^{184}W^{3+}$ STB ×10$^3$ | $^{16}OH_2^+$ MRP ×10$^2$ | $^{16}OH_2^+$ STB ×10$^2$ | $W^{4+}/(W^{3+}+W^{4+})$ CSR×10$^{-2}$ | Field estimate (V/nm) | $H_3O^+/H^+$ ratio |
|---|---|---|---|---|---|---|---|---|---|---|
| Reflectron-fitted | RT | LP | 1.50 | 13.0 | 14 | 9.5 | 0.3 | 0.3 | 48.4 | 0.138 |
| Reflectron-fitted | RT | VP | 0.31 | 12.0 | 11.4 | 9.5 | 6.2 | 9.3 | 55.4 | 0.0480 |
| Straight | RT | LP | 0.50 | 7.9 | 1.6 | 6.1 | 0.2 | 3.6 | 53.2 | 0.0699 |
| Straight | RT | VP | 0.52 | 6.5 | 1.8 | 6.4 | 0.2 | 6.9 | 54.7 | 0.0782 |
| Straight | Cryo-A | LP | 0.38 | 9.0 | 3.8 | 7.2 | 1.5 | 2.0 | 52.0 | 2.10 |
| Straight | Cryo-A | VP | 0.35 | 5.7 | 4.1 | 2.6 | 0.7 | 8.2 | 55.1 | 1.01 |
| Straight | Cryo-B | LP | 0.45 | 8.8 | 2.8 | 2.1 | 0.1 | 5.0 | 53.9 | 0.357 |

**RT: Room temperature - Cryo-A: Cryo-scenario A - Cryo-B: Cryo-scenario B**
**VP: Voltage pulsing - LP: Laser pulsing**

## Discussion

Despite the mass spectra from each sample in this study bearing some obvious differences (as shown in **Figure 5**), they were compared to mass spectra from other, 'graphene-free' studies investigating the field evaporation of pure water or hydrated materials using APT (El-Zoka, et al., 2020; Schwarz, et al., 2020; Schwarz, et al., 2021; Stender, et al., 2022). These studies all involve cryogenically prepared samples with a volume of water considerably larger than that of the samples studied in this work (typically > $1\times10^6$ ions collected). Regardless of whether the water was deuterated or not, several groups of peaks were detected, with an intensity decreasing as the mass-to-charge ratio (m/z) increases. The most prominent peaks in these groups have been identified as $(H_2O)_x(H_3O)^+$ ($0 \leq x \leq 4$) and therefore have a distinctive periodicity of 18 Da (20 Da for deuterated water). While ions stemming from the frozen water with m/z as high as 91 Da were detected by Schwarz, et al. (2020), in this study no peaks higher than 44 Da can be assigned to water-related ions other than the thin layer of $WO_{x,}$ ($x\leq2$). The detection of ions with low m/z suggests the following hypotheses:

1. The detected ions are molecular ions with a high charge-state (m/z).
2. Ions with high m/z were field evaporated with a TOF longer than the duration of a pulse, and therefore not detected or detected on the next pulse window.
3. The ions stemming from the water were evaporated as light, singly charged ions.
4. Any field-evaporated ions with high m/z were dissociated prior to hitting the detector.
5. The water ice is in a vitreous state that field evaporates differently to crystalline ice.

The first hypothesis is less likely due to the relatively low field-conditions improved by the graphene-coating as indicated by the low standing voltage and despite the high field estimated for the pure W substrate. The second hypothesis is unlikely for experiments conducted on a straight FP instrument due to the large m/z window, even at low voltage. Although no evidence supports this hypothesis for the voltage-pulsed experiment conducted on the reflectron-fitted instrument, the broad peaks/tails observed in the lower m/z range for the laser-pulsed experiment mass spectrum (**Figure 5a**) were possibly caused by the "wrap-around" effect described by Larson, et al. (2013), and can be explained by the very low voltage ($\leq 2$ kV) as shown in **Figure 3b**. Furthermore, the mass history in **Figure 3c** shows the evaporation trails resulting from this artefact disappear with increasing ion sequence, which indicates that the m/z window is widening due to the increasing voltage, thus eventually allowing these heavier ions to be properly detected once the necessary voltage is reached. The third hypothesis is supported by the absence of peaks at half (or fractional) Da values and the presence of graphene, which could help identify peaks at 28 and 44 Da as $CO^+$ and $CO_2^+$, respectively. It should also be noted that the graphene-encapsulated water is expected to be in the vitreous state as reported by Zhang, et al. (2022). Although simulation by Segreto, et al. (2022) showed little difference in the field evaporation behaviour of crystalline and vitreous water, the vitreous state of water in this study could explain the highlighted mass spectral differences. The fourth hypothesis could be dismissed due to the low-field conditions caused by the presence of graphene (Exertier, et al., 2022); however, the influence of the high evaporation field W substrate must be considered. As mentioned by Pinkerton, et al. (1999) the electrostatic field required to field evaporate water was found to be dependent on the thickness of the water layer. More particularly, the field required to ionise and field evaporate water was found to decrease with thinner layers of water (Pinkerton, et al., 1999). Since water is already a low-field material that has been reported to field evaporate as clusters with high m/z (Panitz & Stintz, 1991; Pinkerton, et al., 1999), the dissociation of water clusters appears as a likely explanation to the low m/z peaks detected for all scenarios in this study. Whilst a dissociation analysis could help understand the actual mechanisms involved here (Saxey, 2011; Zanuttini, et al., 2017), the low water ion counts in the different datasets and the kinetic energy compensation of the reflectron-fitted instrument make the analysis of molecular dissociation challenging (Saxey, 2011). Nonetheless, correlation histograms were produced for each dataset and presented in Section 2 of the supplementary materials. Minimal new information was obtained from these correlated evaporation histograms. **Figure S6** and **Figure S10** show the dissociation of W oxide ions into W and O ions ($WO^{3+} \rightarrow W^{2+} + O^{+}$) for the samples analysed on the straight flight path instrument using laser pulsing, at room temperature and following cryo-scenario B, respectively. Furthermore, the $H_3O^+$ (19 Da) to $H^+$ (1 Da) ratio was calculated for each experiment and shown in **Table 3** and **Figure S3** to help investigate the molecular dissociation phenomenon. **Figure S3** shows no obvious correlation between this ratio and the field estimate, however the low values are consistent with a higher field, especially for the voltage-pulsed experiments, which would also explain the absence of $H_3O^+$ peak in the dataset acquired by voltage-pulsing on the reflectron-fitted instrument. Finally, the fifth hypothesis (that the data in this work is representative of vitreous ice, whilst other studies represents crystalline ice) is supported by Zhang, et al. (2022), however no further studies have been conducted on the difference in the field evaporation of vitreous versus crystalline ice.

As noted previously, the current work studies only small volumes of water encapsulated between the graphene foil and the W substrate. Although a previous study estimated the volume of encapsulated water prior to field evaporation to be in the range of ~ $10^{-18}$ L to $10^{-21}$ L (Qiu, et al., 2020a), the size of the datasets (i.e., the ionic count and reconstructed data volume) studied in this work as well as in other related studies (Qiu, et al., 2020b; 2020c; 2020d; Zhang, et al., 2022) do not reflect this volume range. The volume corresponding to the detected signal stemming from water is thought to be much higher than what is actually measured, particularly because of nano-fractures evidenced by the relatively unstable detection rate measured during the field evaporation of graphene-encapsulated water shown in **Figure 4**, for instance, and in **Figure S1** regardless of the pulsing mode and instrument flight path. These nano-fracture events also partly explain the higher background level. As an example, using the density of water at room temperature (as a proxy for amorphous ice), its molar mass and Avogadro's number, a 1 nm thick layer of water with a radius of 50 nm holds $2.6\times10^5$ water molecules. In terms of an APT data scenario, considering a 50% detector efficiency and assuming all the water molecules from this hypothetical layer were all evaporated individually without nano-fractures, there would be half this number of water molecules. Given the typical thickness of water that is graphene-encapsulated several times on the tip of an APT needle is typically around 10 nm (Qiu, et al., 2020b; 2020c; 2020d), and that an APT needle usually has a radius of curvature greater than 50 nm (Gault, et al., 2012), a graphene-encapsulated water APT dataset should have at least around $1\times10^6$ detected water ions (i.e., half the number of molecules described above, times ten). In this study and in previous studies (Qiu, et al., 2020b; 2020c; 2020d), graphene-encapsulated water datasets typically have between $1\times10^4$ and $1\times10^5$ water ions with little to no large water clusters, which supports the postulate that a large amount of water is lost during these nano-fractures. Whilst it is possible that some of the signal is lost due to potential molecular dissociation of $H_3O^+$ into $H^+$ and neutral $H_2O$ ($H_3O^+ \rightarrow H^+ + H_2O$), this alone cannot explain the small dataset size.

For other studies where large volumes of water were field-evaporated using laser pulsed APT, the apex of the tip was located far from the metal substrate (70 to 100 μm as reported by Schwarz, et al. (2020)), thus, preventing the metal substrate from dissipating the heat from the laser and causing thermal tails in the mass spectra (Segreto, et al., 2022). Here, the absence of thermal tails in laser-pulsed mass spectra from graphene-encapsulated specimens prepared at room temperature (**Figure 5a** and **Figure 5c**) suggests the combined influence of the graphene layer and W substrate help dissipate the heat. Whilst the properties of thermally conductive substrates such as W have been used to dissipate the heat in previous laser-pulsed APT studies, for instance by increasing the shank angle of the needle specimens (Gault, et al., 2012), in this case heat dissipation is more likely caused by the graphene layers sandwiching the water layers due to the extremely high thermal conductivity of graphene, even at cryogenic temperature (Balandin, 2011). Moreover, for the sample whose graphene layer was removed using cryo-FIB milling, the mass spectrum (**Figure 5g**) has the lowest MRP of all mass spectra for the peaks corresponding to $^{16}OH_2^+$ and $^{184}W^{3+}$. These observations suggest that small volumes of liquid that have been field evaporated in the presence of graphene, have an increased mass spectral quality compared to graphene-free samples due to the high thermal conductivity of the graphene. This highlights an advantage of using graphene encapsulation in addition to the fact that graphene-encapsulated water samples are vitreous upon contact freezing with a cryogenic stage (Zhang, et al., 2022). In the context of future studies of hydrated biological materials where vitreous ice is required to preserve the structure of the biological sample, high pressure freezing may be necessary to achieve vitreous ice formation for large samples (i.e., larger than several hundreds of nanometres).

All the datasets presented here were prepared following the same protocol involving a pre-run prior to graphene coating. However, as opposed to previous research (Exertier, et al., 2022) where tip shape consistency was essential to measure field-related components, here the main goal of the pre-run was to clean the tips of any impurities, and therefore the pre-runs were interrupted at different voltages from one sample to another. Here, a larger number of water-related ions was observed for samples with a smaller radius at the apex, which can easily be visualised using mass-to-charge and voltage history diagrams as shown in **Figure 3**. Due to the important evaporation field difference between water and tungsten, we postulate that reducing the initial radius of the tungsten substrate needle would enable the tungsten substrate to start field evaporating at a lower voltage, which in turn would reduce the stress on the sample, thus reducing the chance of data loss due to nano-fractures. Based on these observations, we recommend the use of initial substrate needles with a smaller radius or with a low-field material to maximise the sample yield.

Several data quality metrics have been compared for instruments with different flight paths. As expected, the MRP is higher for the reflectron-fitted instrument than for the straight flight path instrument. However, there is a major trade-off with the increased mass spectral quality: a longer flight path causes a reduced TOF window for ions to reach the detector before the next pulse, which may cause a "wrap-around" effect resulting in loss of data from heavy molecular ions, especially at low voltages. Similarly, whilst a higher pulse frequency usually reduces the background level, it also contributes to reducing the detectable m/z range. Therefore, in the case where large molecular ions are expected, it may be preferable to use a reduced pulse frequency. The two instruments used in this work have different detector efficiencies, i.e., 57 % for the straight flight path instrument and 50 % for the reflectron-fitted instrument, however no notable interferences were observed in the datasets.

## Conclusions

In this work, graphene-encapsulated vitreous water ice was analysed using APT with different acquisition conditions. The water signals from the different acquisition scenarios were investigated, then compared to each other and to other studies on APT of pure water ice prepared by droplet freezing followed by cryo-FIB milling in the literature. The findings of these analyses are summarised as follows:

- Fewer molecular ions of large m/z were detected (only up to 45 Da) compared to other studies involving larger volumes of water (molecular ions with m/z up to ~100 Da).
- Thermal tails were reduced in the presence of graphene for laser-pulsed experiments, both in this work (i.e., with/without graphene) and compared to that in the literature with large volumes of frozen water.

- For most experimental scenarios, the composition of graphene-encapsulated water consisted of at least $OH_x^+$ ($0 \leq x \leq 3$), as well as $C^+$, $CO^+$ and $COH^+$. This compositional basis can be used in further studies involving the graphene-encapsulation of other materials.
- Cryo-specimen preparation has been attempted, but frost build-up was not easily controlled with the employed method and instruments.

Analysis of data quality metrics revealed several APT parameter optimisation pathways:

- A smaller radius at the specimen tip apex of the initial metal substrate needle improves the yield and increases the detected volume of water.
- Due to the low electrostatic field conditions, a low pulse frequency is required to avoid the "wrap around" effect, especially with laser-pulsing on a reflectron-fitted flight path instrument which offers the best mass resolving power.

Out of the few sample preparation methods enabling APT analysis of pure water, graphene-encapsulation involves the smallest water volumes. This can have benefits when analysing nanoscale biological samples such as individual proteins, which can be successively sandwiched in thin vitreous ice layers. Cryo-specimen preparation has been conducted; however further improvements are necessary to ensure the specimens remain free from frost-induced damage. Improvements to cryo-instrument designs would allow for better control of the humidity level (glovebox) and simpler transfer processes to avoid potential contamination and therefore ensure better cryo-protocols and more accurate APT analyses of hydrated materials.

## Acknowledgements

This study was funded by the Australian Research Council (DP180103955). The authors acknowledge the instruments and scientific and technical assistance of Deakin University's

Advanced Characterisation Facility, a Linked Laboratory of Microscopy Australia, and Sydney Microscopy and Microanalysis, the Sydney node of Microscopy Australia. The authors thank Dr. Limei Yang for her help with cryo-FIB APT sample preparation.

# Supplementary Materials

## Section 1: 3D reconstructions and selected evaporation data

The data from all the samples analysed in this study are summarized in this section. **Figure S1** shows for each sample a 3D reconstruction generated using the "Tip profile" method, as well as the evolution of the voltage, the detection rate, and the mass-to-charge ratio of the evaporated ions plotted against the evaporated ion sequence. **Figure S1** is split in four panels, each comparing the results for samples analysed using voltage pulsing and laser pulsing according to a specific scenario. **Figure S1.1** and **Figure S1.2** show the results from graphene-encapsulated samples prepared at room temperature and analysed using a reflectron-fitted instrument and a straight flight path instrument, respectively. **Figure S1.3** shows the results from samples that were graphene-encapsulated at room temperature, then plunge-frozen and cryogenically transferred to the straight flight path instrument (cryo-scenario A). **Figure S1.4** shows the results from a sample that was graphene-encapsulated at room temperature, plunge-frozen and cryogenically transferred to a cryo-FIB/SEM instrument to undergo cryo-FIB milling, then cryogenically transferred to the straight flight path instrument (cryo-scenario B).

It is noted that the data shown in **Figure 3** and **Figure 4** were reproduced in **Figure S1** (columns a and f, respectively) to allow for an easier comparison with the other samples.

| Reflectron | |
|---|---|
| RT specimen preparation | |
| **Laser pulsing** | **Voltage pulsing** |

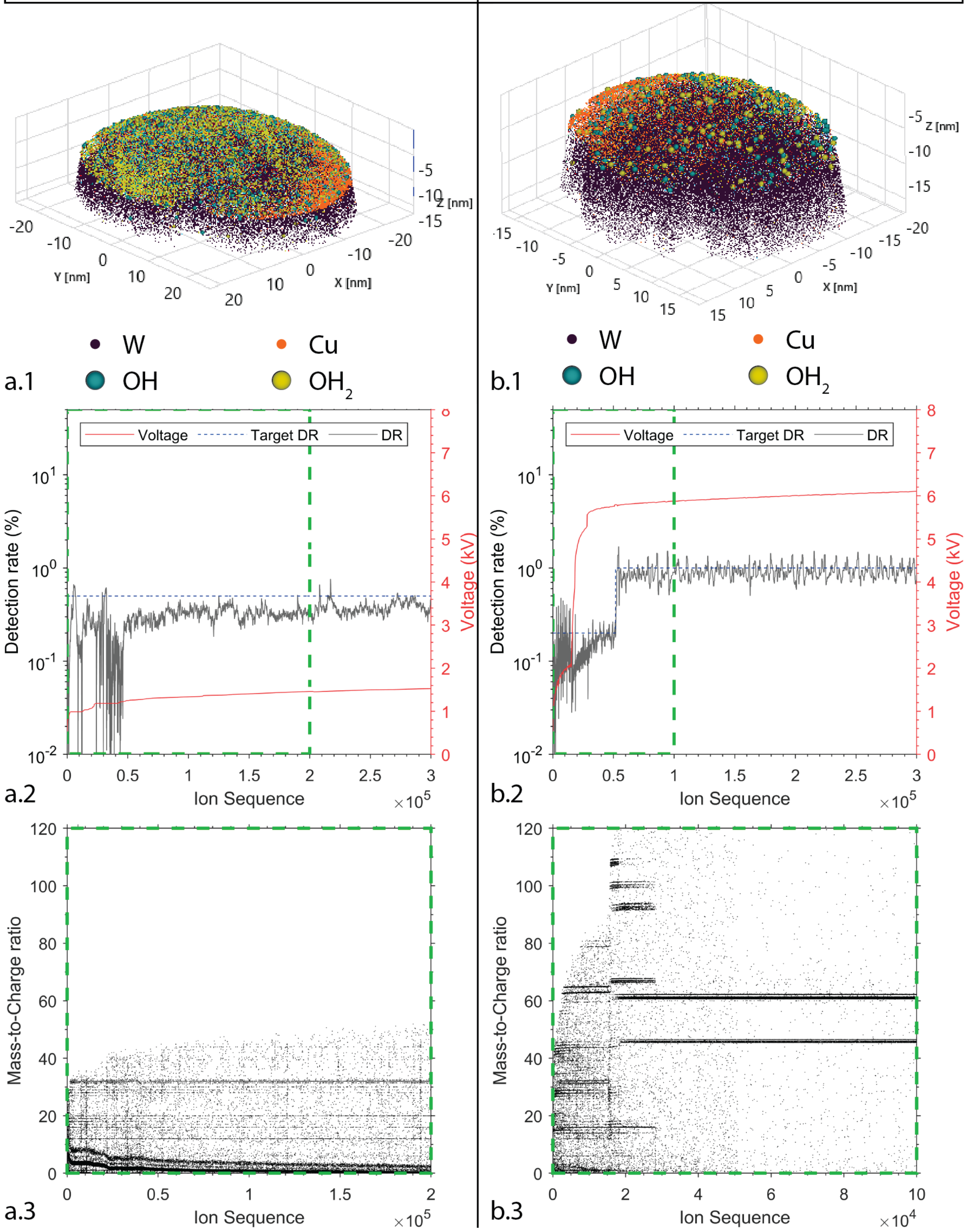

**Figure S1.1** Samples prepared at room temperature analysed using a reflectron fitted instrument.

| Straight Flight Path | |
|---|---|
| RT specimen preparation | |
| **Laser pulsing** | **Voltage pulsing** |

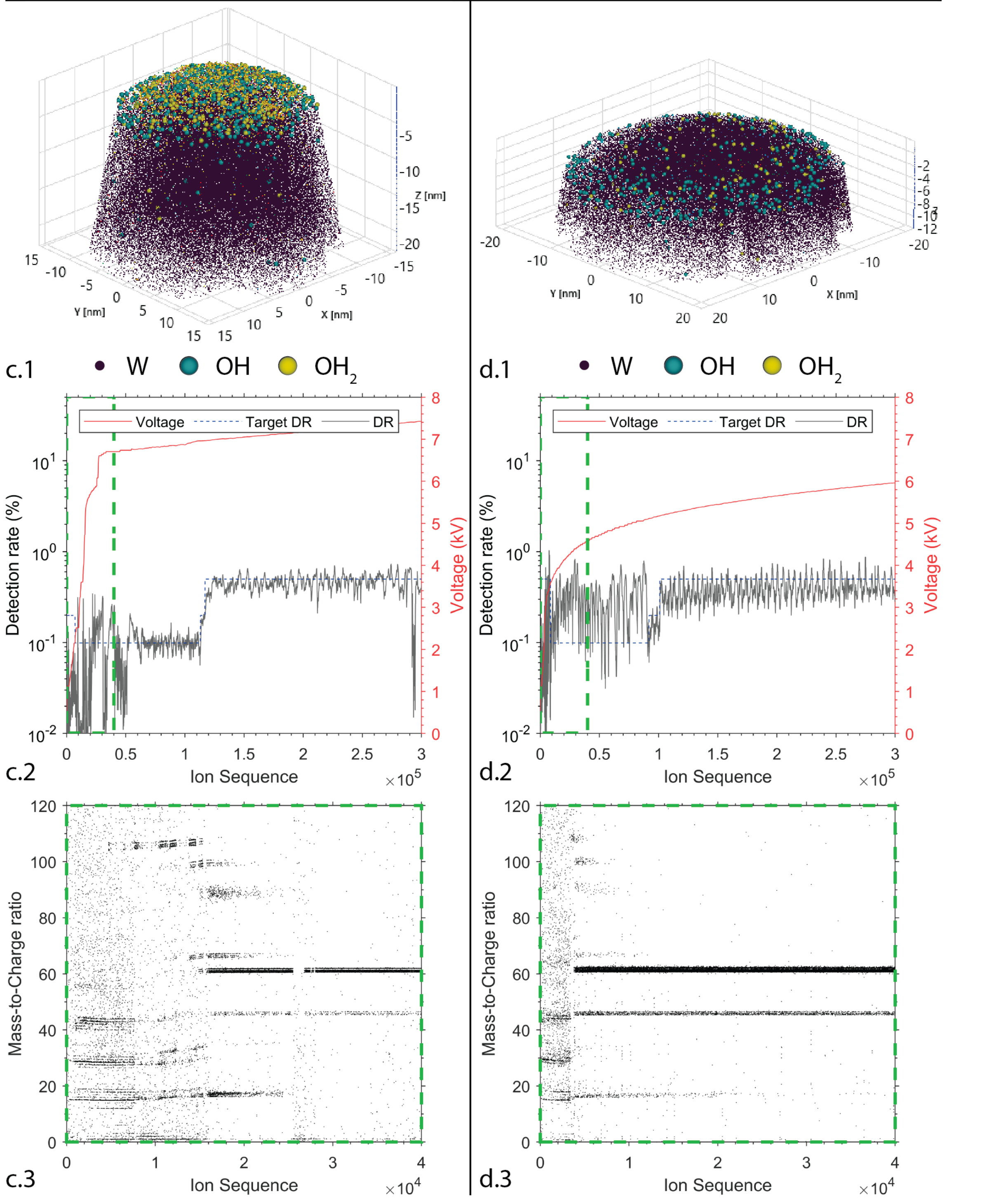

**Figure S1.2** Samples prepared at room temperature analysed using a straight flight path instrument.

| Straight Flight Path | |
|---|---|
| Cryo specimen preparation - Scenario A Cryo-A (No FIB) | |
| **Laser pulsing** | **Voltage pulsing** |

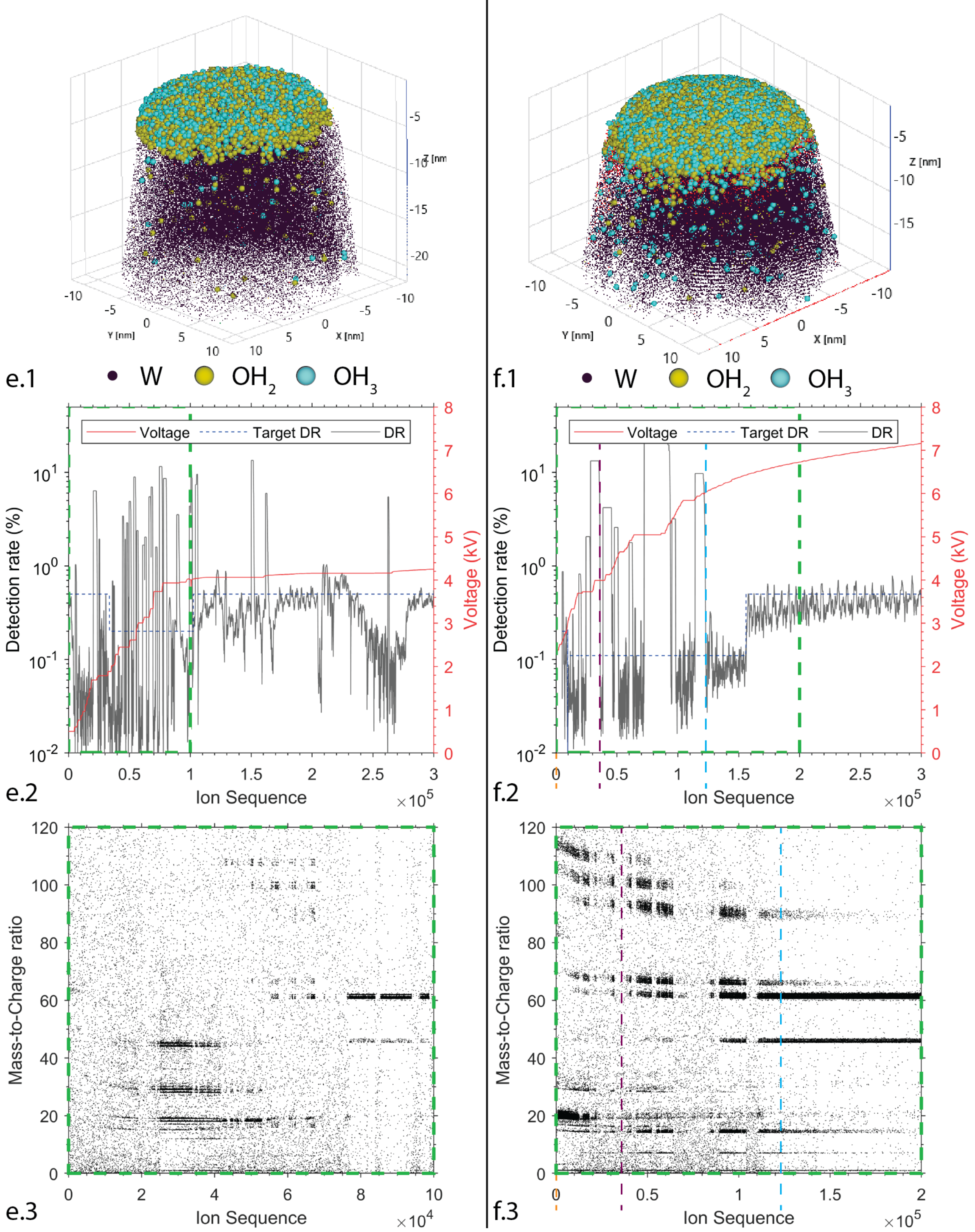

**Figure S1.3** Samples prepared following cryo-scenario A analysed using a straight flight-path instrument.

| Straight Flight Path |
| --- |
| Cryo specimen preparation - Scenario B (FIB) |
| **Laser pulsing** |

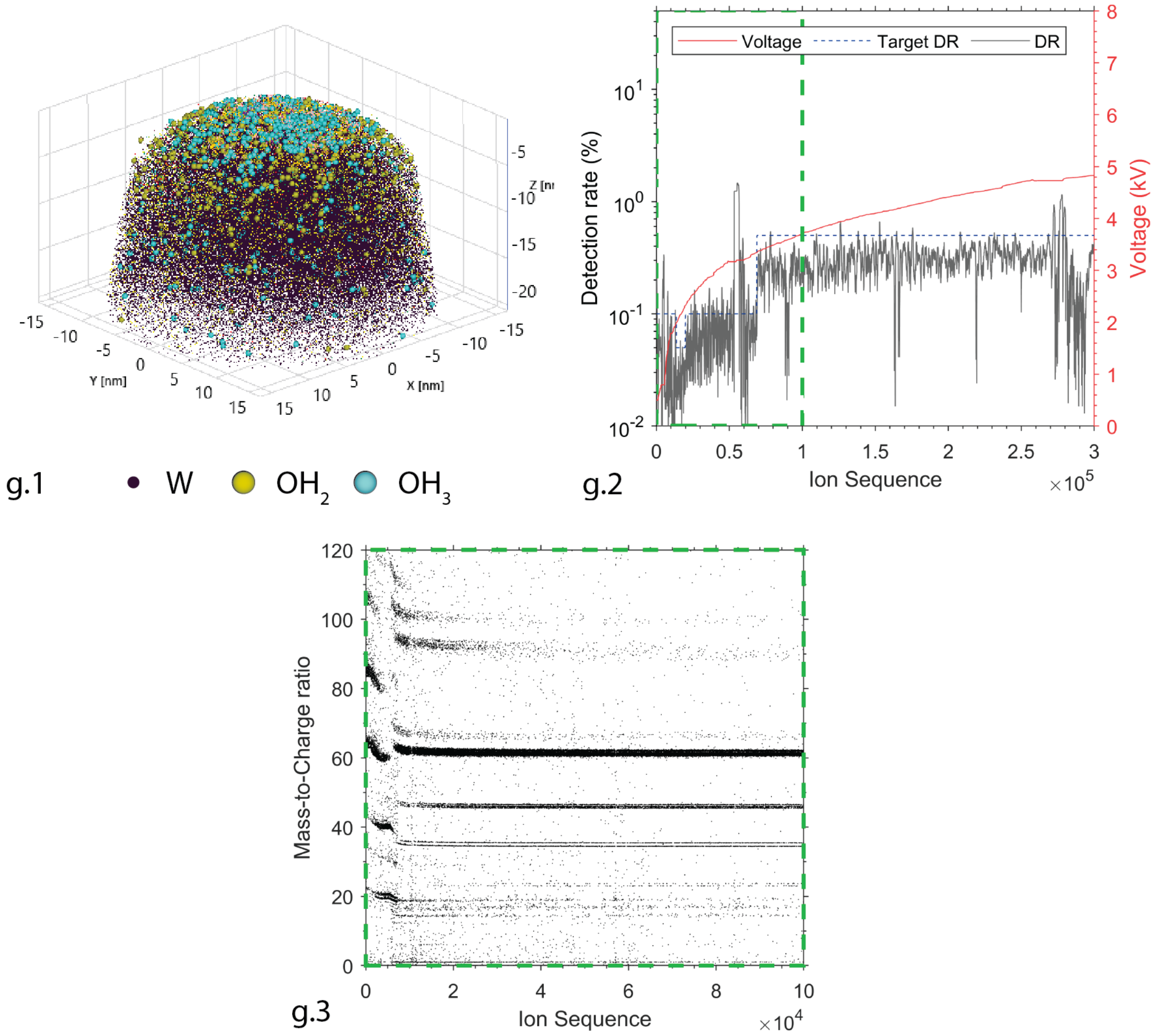

**Figure S1.4** Sample prepared following cryo-scenario B analysed using a straight flight-path instrument.

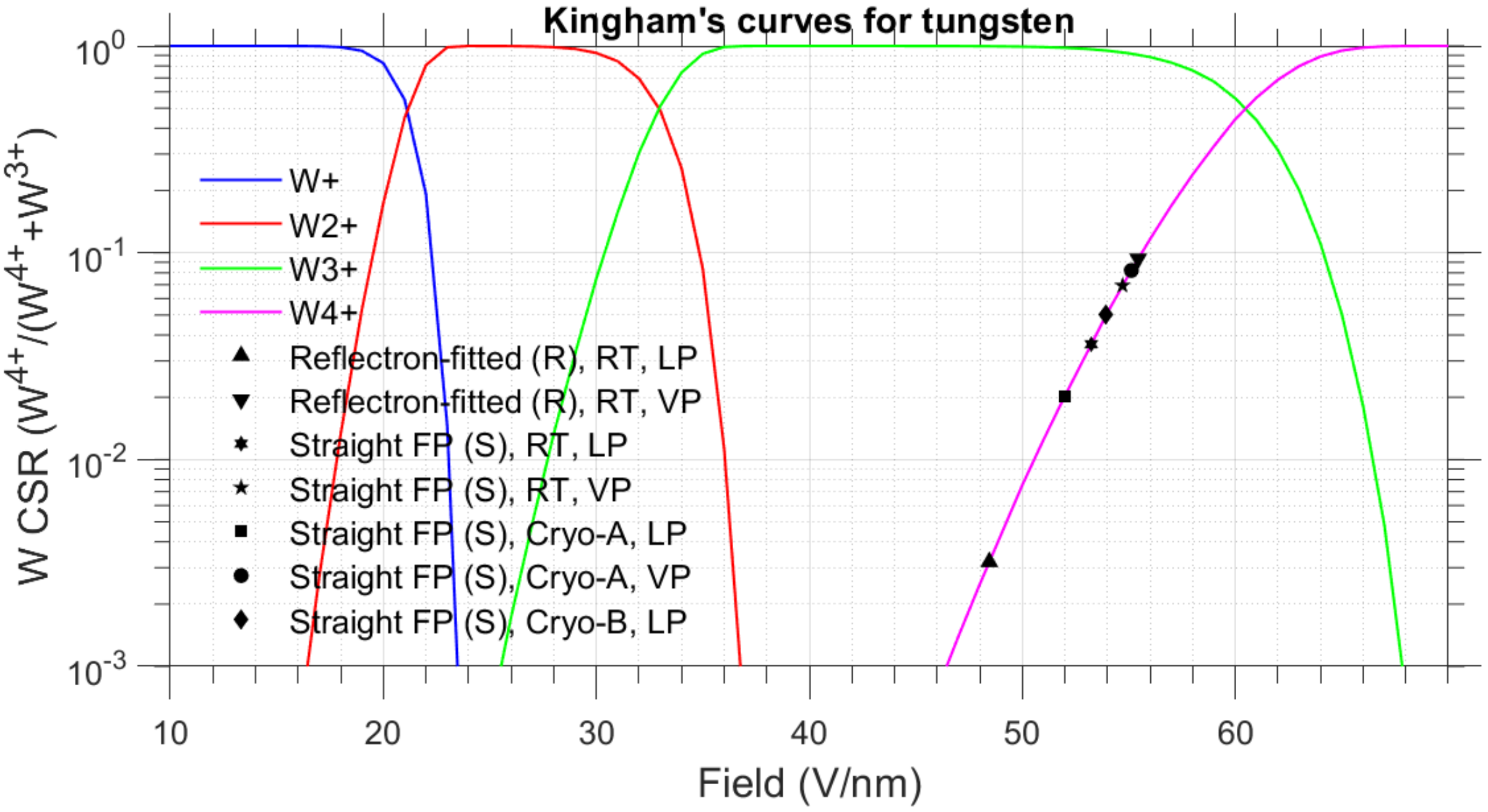

**Figure S2** Kingham's curves calculated for W showing the charge-state-ratios against the electrostatic field estimate. The $W^{4+}/(W^{4+}+W^{3+})$ ratios calculated for each specimen are also shown.

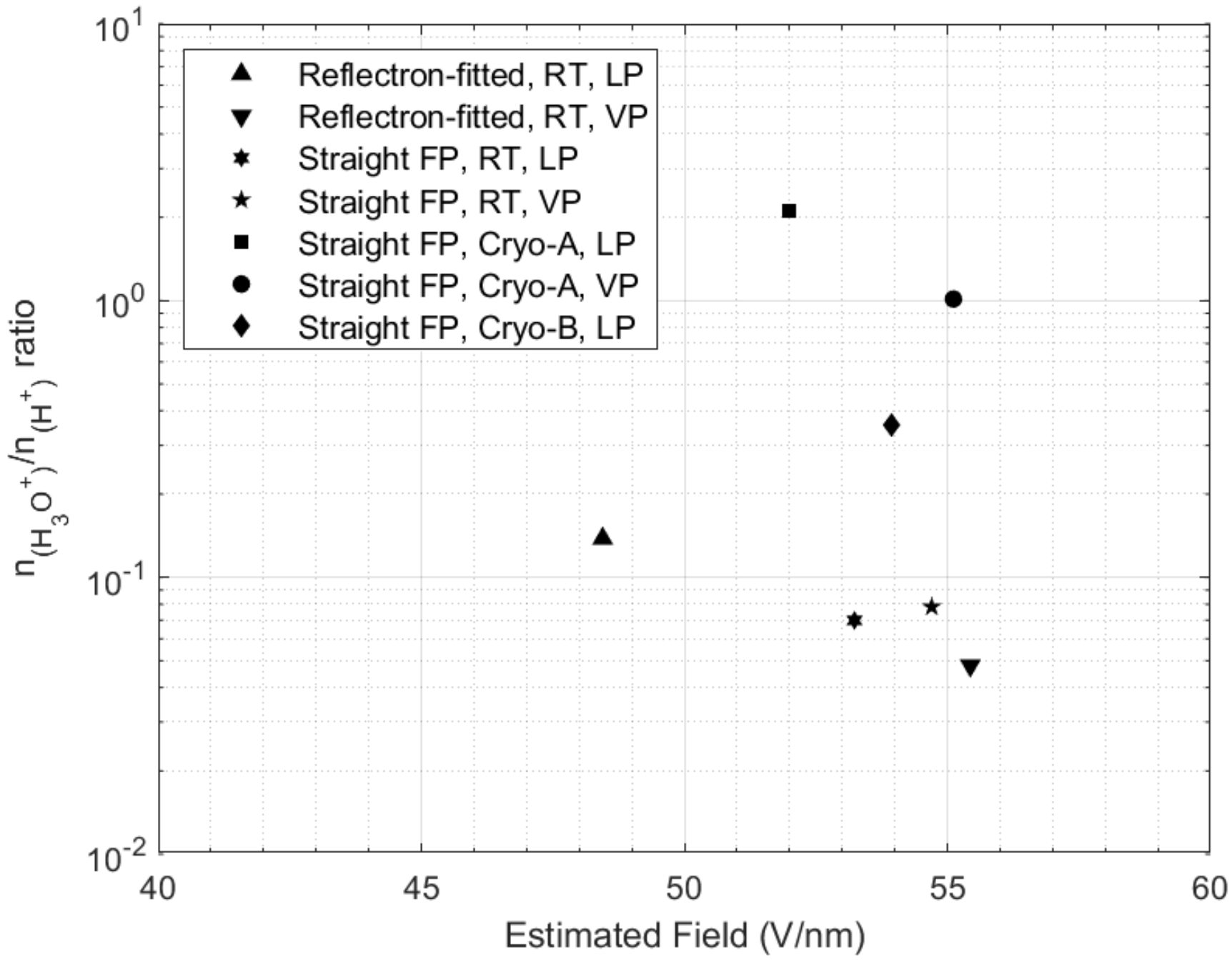

**Figure S3** Ratio of the $H_3O^+$ concentration to $H^+$ concentration plotted against the electrostatic field estimated using **Figure S2**.

## Section 2: Correlated evaporation histograms for each dataset

Correlation histograms as defined by Saxey (2011) are presented in this section. As explained by Di Russo, et al. (2020), such analysis is futile for datasets obtained using a reflectron-fitted instrument, due to the energy compensating effects of the instrument. Correlated histograms are presented for pure water samples prepared at room temperature and analysed using a reflectron-fitted instrument nonetheless (**Figure S4** and **Figure S5**). The following figures correspond to pure water prepared at room temperature and analysed using a straight flight path instrument (**Figure S6** and **Figure S7**), as well as the pure water samples prepared using the cryogenic temperature preparation scenarios cryo-A and Cryo-B (**Figure S8**, **Figure S9**, and **Figure S10**).

Whilst little information is shown by these correlated evaporation histograms, the following dissociation can be observed in **Figure S6** and **Figure S10**.

$$\underset{(66 - 67.3\ \text{Da})}{^{182\text{-}186}WO^{3+}} \rightarrow \underset{(91 - 93\ \text{Da})}{^{182\text{-}186}W^{2+}} + \underset{(16\ \text{Da})}{^{16}O^{+}} \quad (1)$$

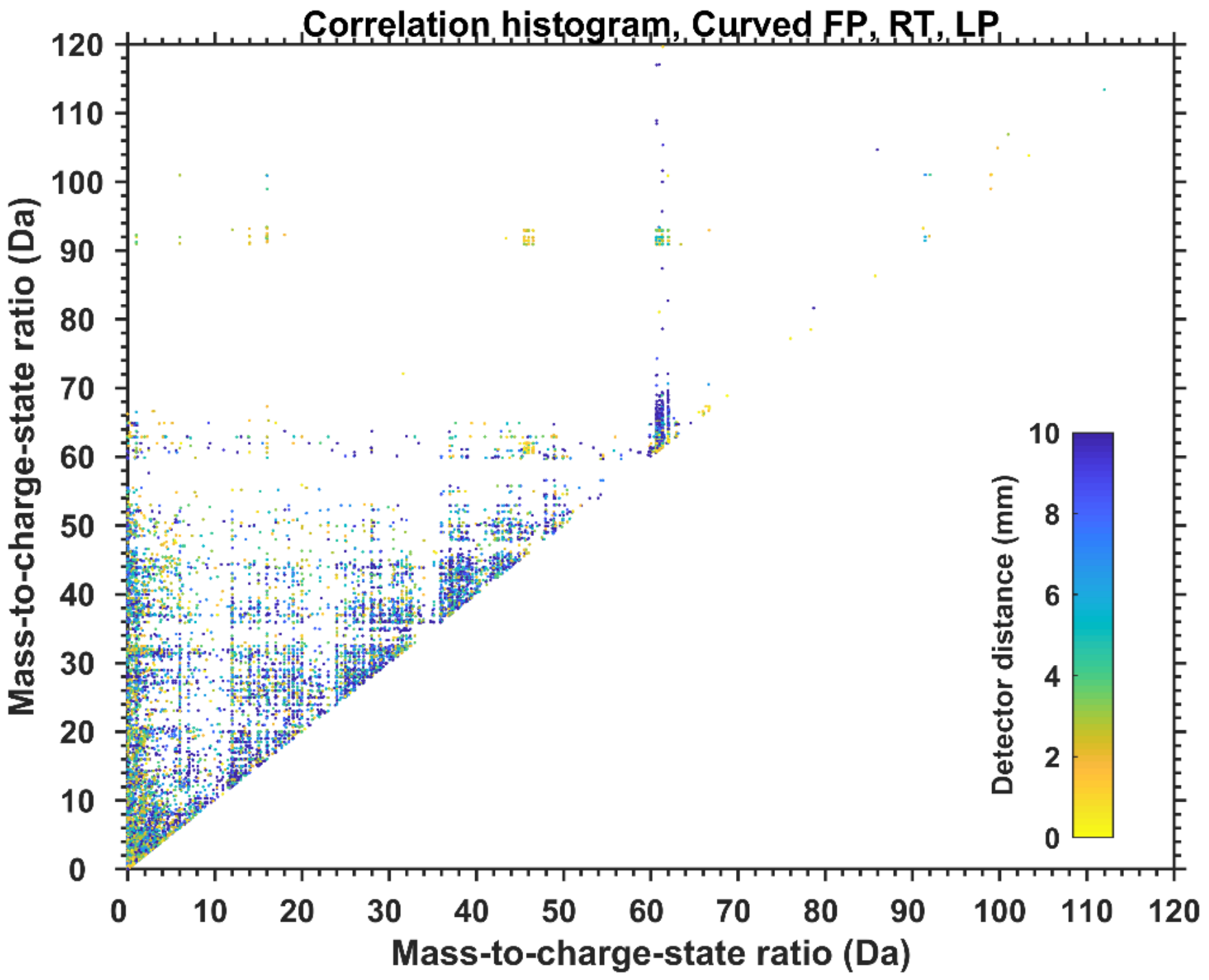

**Figure S4.** Correlation histogram for the water sample prepared at room temperature then analysed using a reflectron-fitted instrument, using laser pulses.

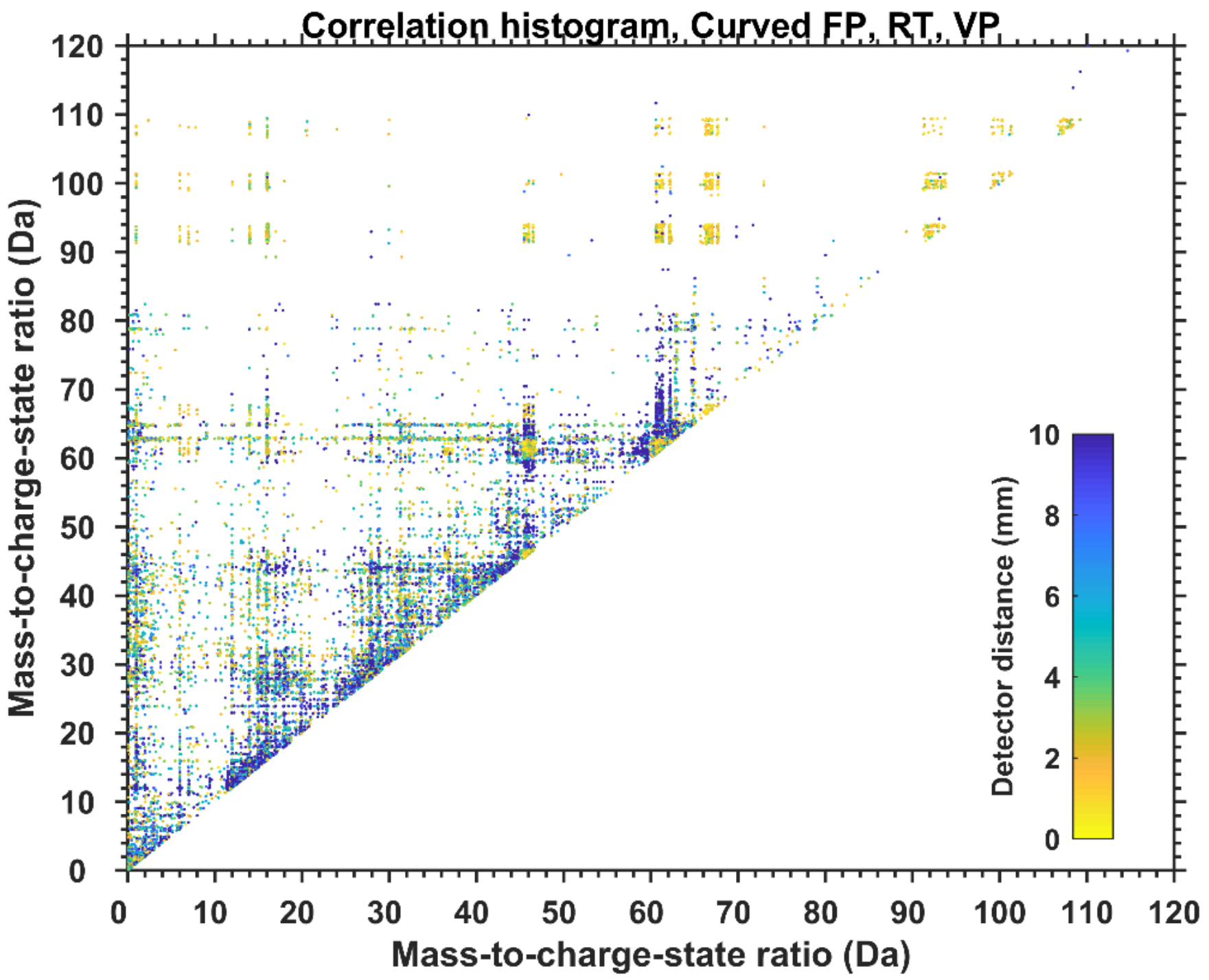

**Figure S5.** Correlation histogram for the water sample prepared at room temperature then analysed using a reflectron-fitted instrument, using voltage pulses.

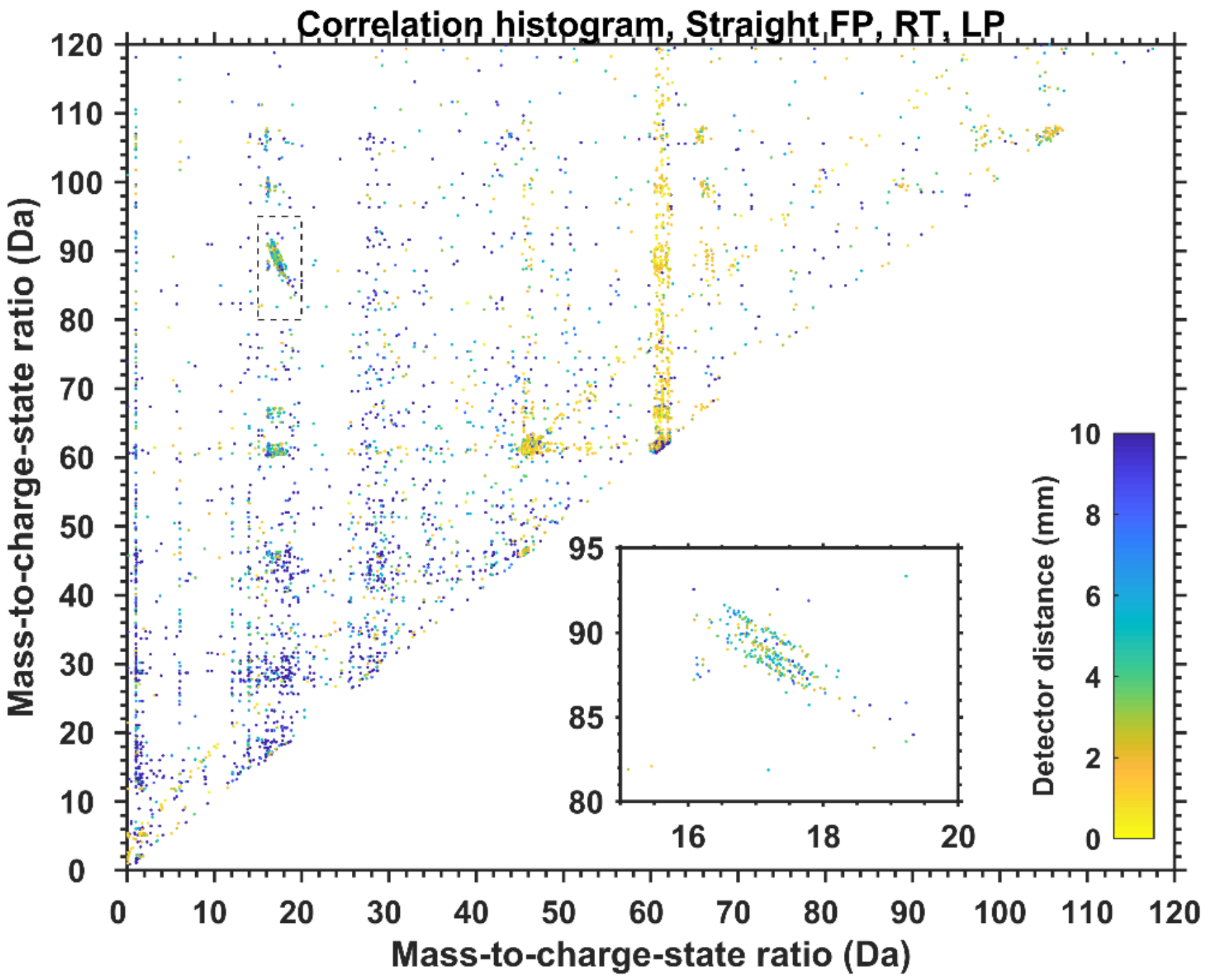

**Figure S6.** Correlation histogram for the water sample prepared at room temperature then analysed using a straight flight path instrument, using laser pulses. Inset shows a dissociation track corresponding to $WO^{3+} \rightarrow W^{2+} + O^{+}$.

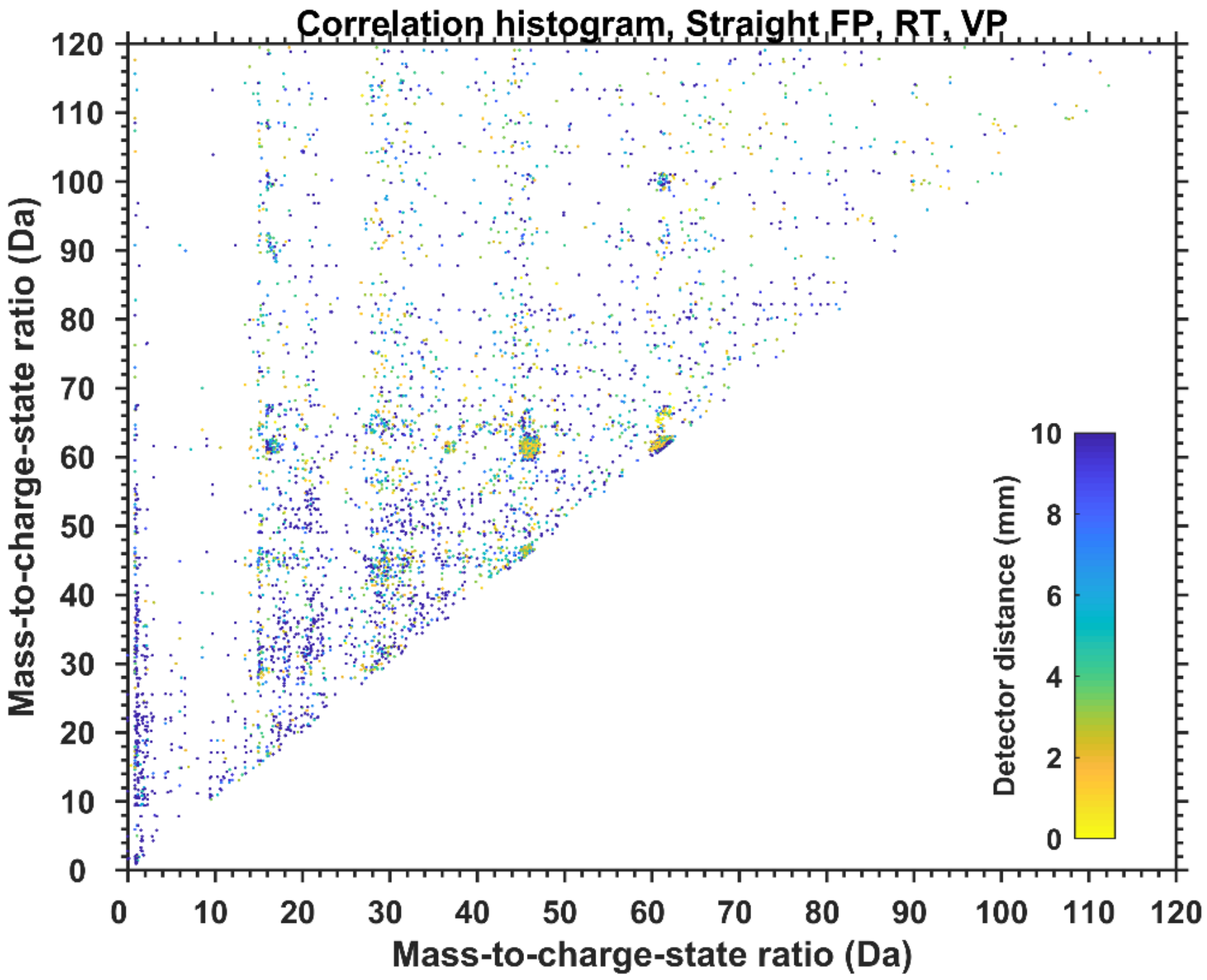

**Figure S7.** Correlation histogram for the water sample prepared at room temperature then analysed using a straight flight path instrument, using voltage pulses.

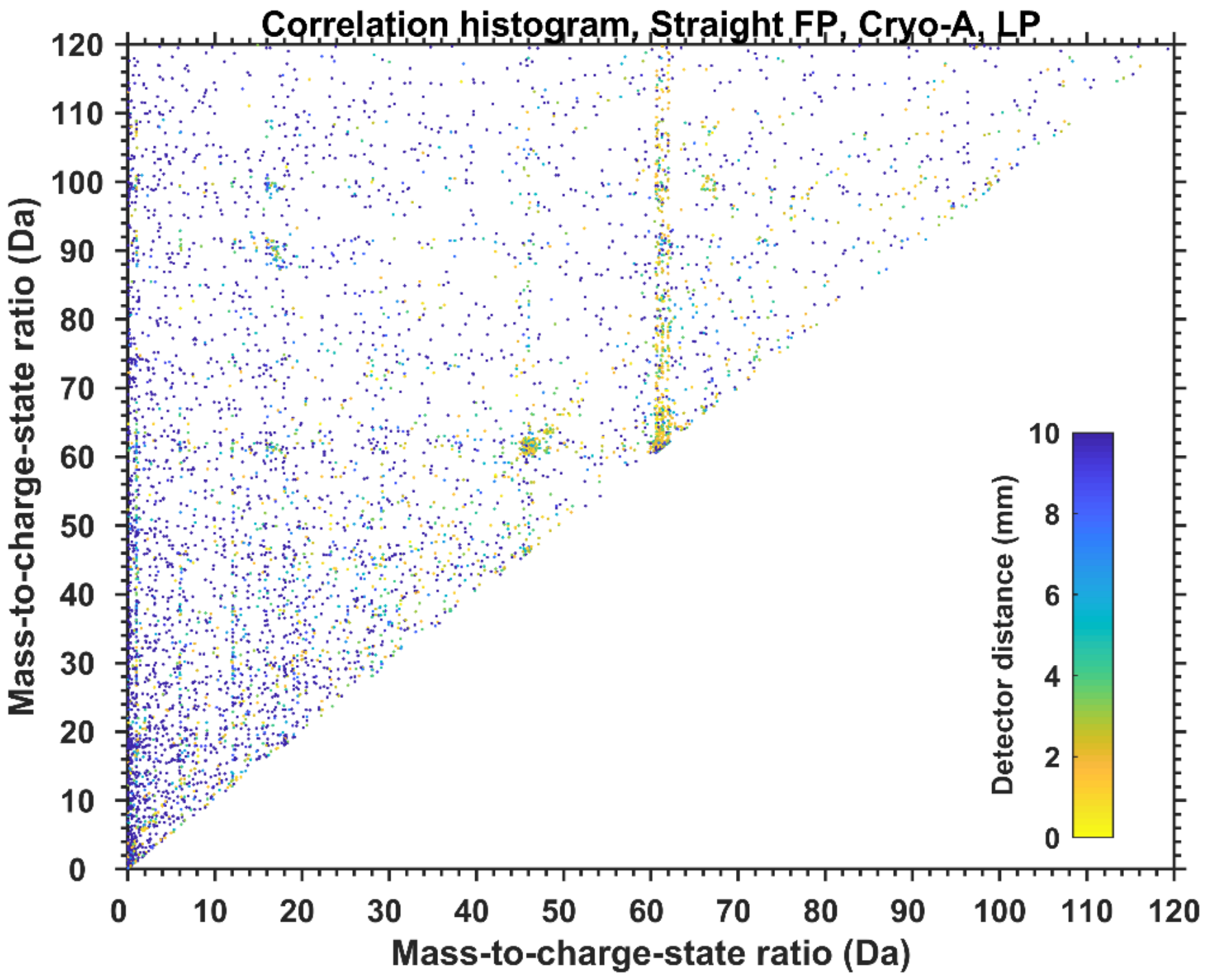

**Figure S8.** Correlation histogram for the water sample prepared at cryogenic temperature (Cryo-scenario A) then analysed using a straight flight path instrument, using laser pulses.

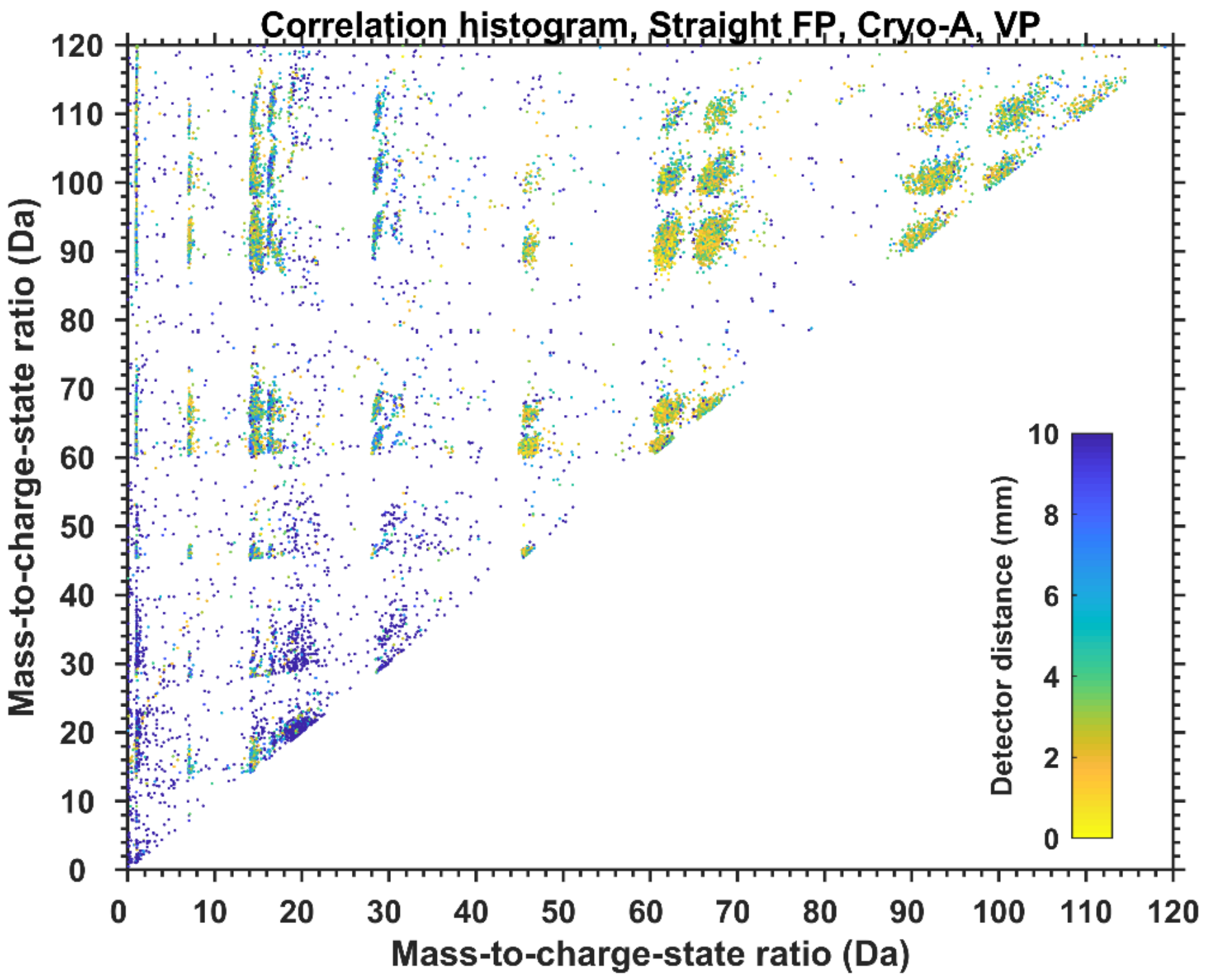

**Figure S9.** Correlation histogram for the water sample prepared at cryogenic temperature (Cryo-scenario A) then analysed using a straight flight path instrument, using voltage pulses.

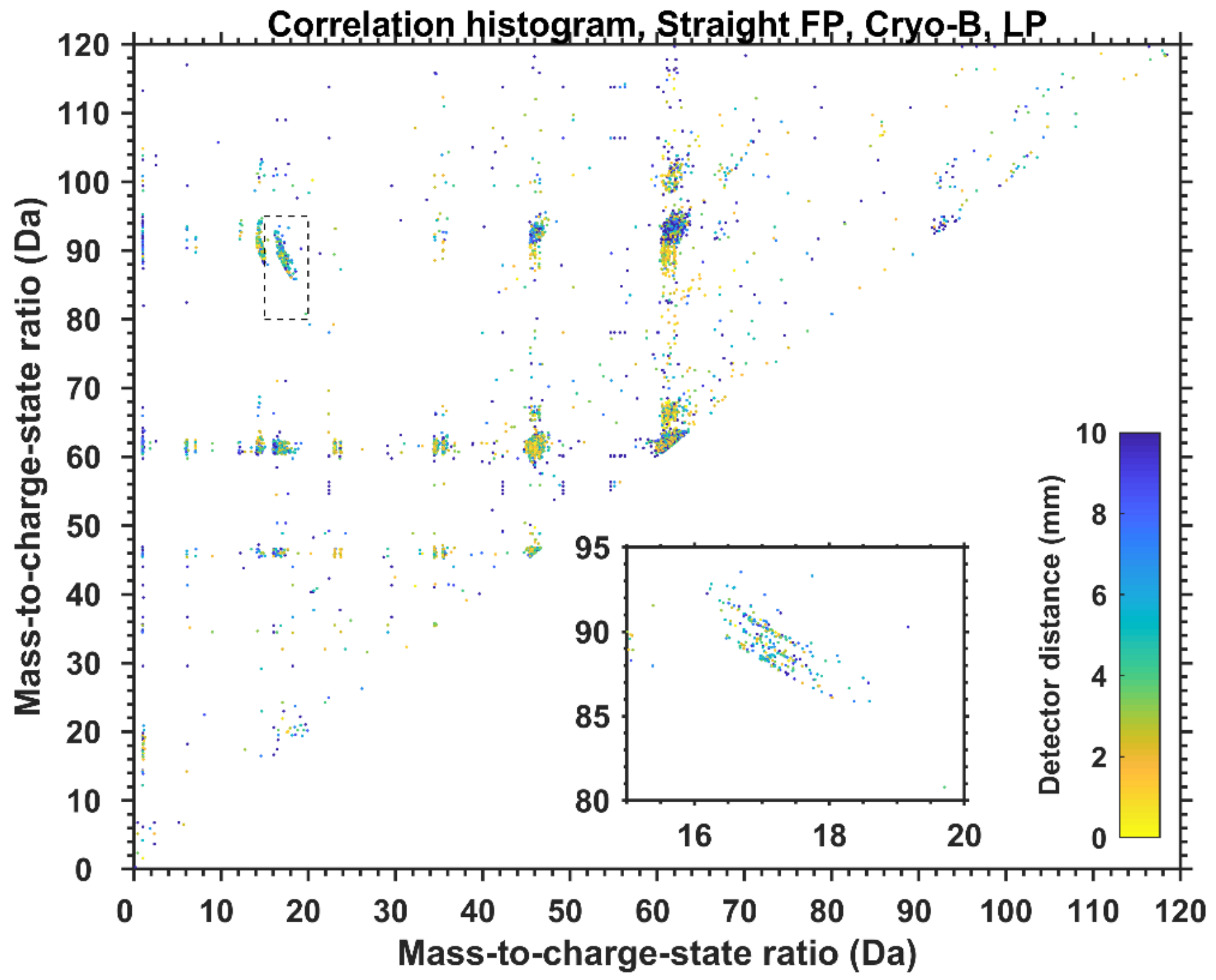

**Figure S10.** Correlation histogram for the water sample prepared at cryogenic temperature (Cryo-scenario B) then analysed using a straight flight path instrument, using laser pulses. Inset shows a dissociation track corresponding to $WO^{3+} \rightarrow W^{2+} + O^{+}$.

## Section 3: Cryo workflow

In this section, the cryo-APT equipment and associated protocols are presented in detail.

The cryo-APT experiments were conducted in Sydney using the workflow summarised in **Figure 1**, split into two scenarios. In scenario A, graphene-encapsulated needles were inserted into a glovebox, then plunge-frozen using liquid nitrogen ($LN_2$). Using a vacuum cryo transfer (VCT) suitcase (Ferrovac, GmbH), the samples were then transferred to the atom probe for analysis. In scenario B, the samples were plunge-frozen in the same fashion. However, prior to loading into the atom probe, the samples were cryo-transferred to a cryo-FIB to undergo cryo-annular milling, to remove the graphene layer so that only pure water remains at the tip of the specimen needle.

The equipment used to enable these cryo-workflows is described by Cairney, et al. (2019) and represented in **Figure S11** where the link between each instrument is the vacuum cryo-transfer suitcase initially presented by Ulfig, et al. (2017).

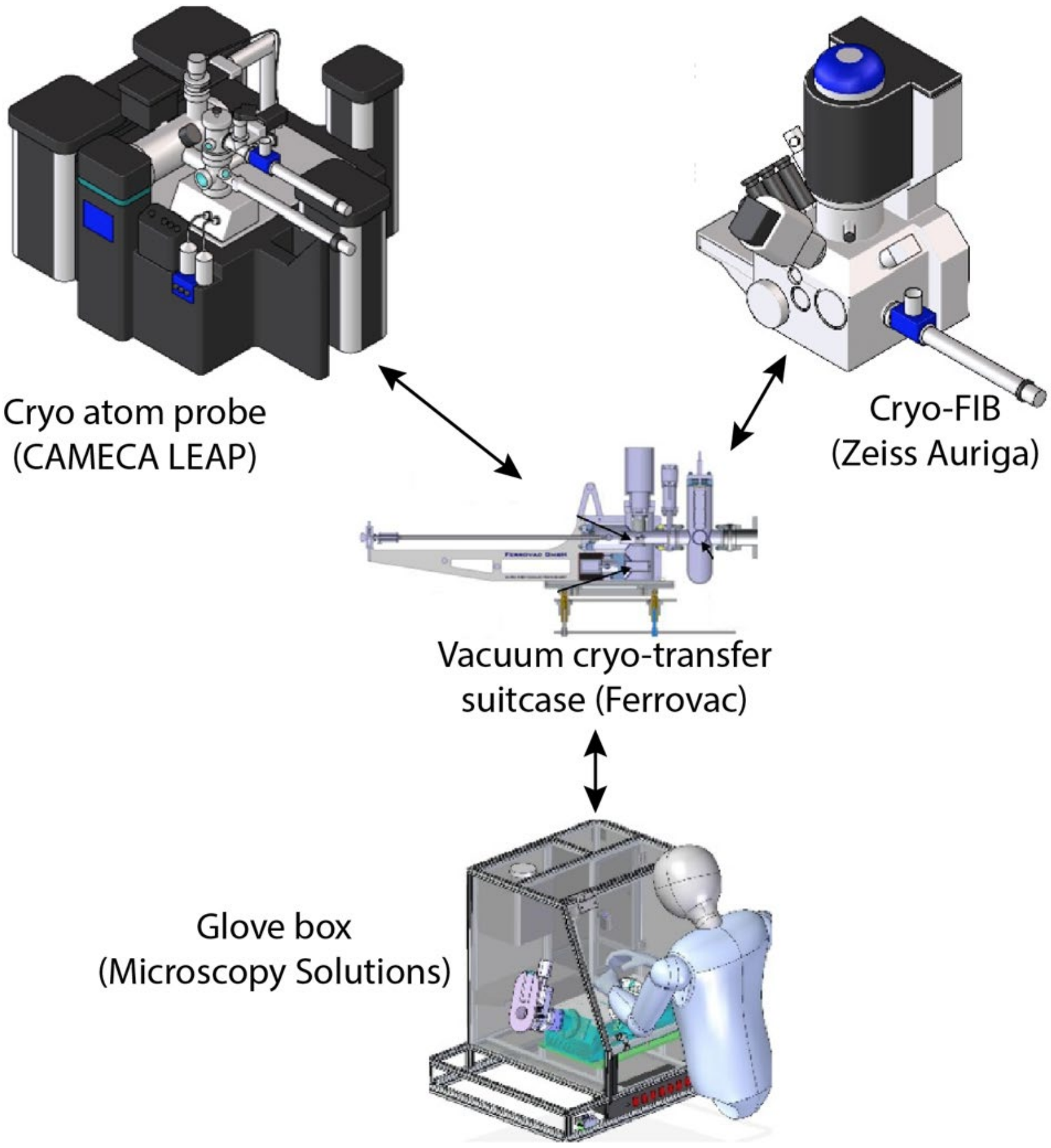

**Figure S11.** Schematics of recently developed solutions enabling the handling of cryo prepared samples. Adapted from Cairney, et al. (2019) and Ulfig, et al. (2017).

## Vacuum cryogenic transfer suitcase

The Ferrovac VCT suitcase is designed around an ultra-high vacuum (UHV) chamber which may be cooled down to cryogenic temperature by an adjacent liquid nitrogen dewar. The vacuum inside the chamber is maintained by an ion pump and can reach pressures as low as $10^{-10}$ mbar (GmbH, 2019), although in practice the vacuum of this particular suitcase during use is around $10^{-9}$ mbar. The UHV chamber is opened by operating a manual gate valve to let samples in and out. The samples are handled by a transfer rod designed to handle Cameca LEAP sample holders (pucks) (Ulfig, et al., 2017). The outer extremity of the transfer rod is equipped with a knob to align it with the connecting interface when transferring a sample in or out. To improve the pumping capacity of the UHV device, the chamber and connecting parts have a small volume, which requires APT specimen needles not be taller than 18 mm, so they do not hit either inner part of the instrument during transfer.

To preserve a good vacuum quality inside the UHV chamber, it is preferable to connect the suitcase to another vacuum chamber, at least HV grade, preferably UHV grade. Prior to connecting the suitcase to another instrument, it may be advisable to blow some nitrogen gas on the inner walls of the exposed part of the other instrument to avoid contamination with dust particles once both sides are connected and open. This suitcase is configured with a KF40 flange, and may be connected to other intruments using either CF or KF flanges. Once the flange is properly tightened, the gate valve on the other instrument (e.g., glovebox, FIB/SEM, LEAP) needs to be opened to equalise the pressure in the gap between the gate valves of each instrument.

When the sample is ready to be transferred in or out of the suitcase, all the remaining gate valves are to be opened, with that of the suitcase being opened last. Then the sample is to be transferred as quickly as possible, following which the gate valve is to be shut in order to maintain high vacuum inside its chamber.

## Glovebox

### Description of apparatus

The glovebox is equipped with 5 main features as labelled in **Figure S12a**: liquid nitrogen supply via top dewar (A), gaseous nitrogen supply (hidden, supplied by nitrogen bottle on **Figure S12b**), an interlock to transfer items in and out of the glovebox (B), several storing shelves (C), a stainless-steel high vacuum interlock (HVI) chamber (D), and a liquid nitrogen bath. Each of these main features is described below.

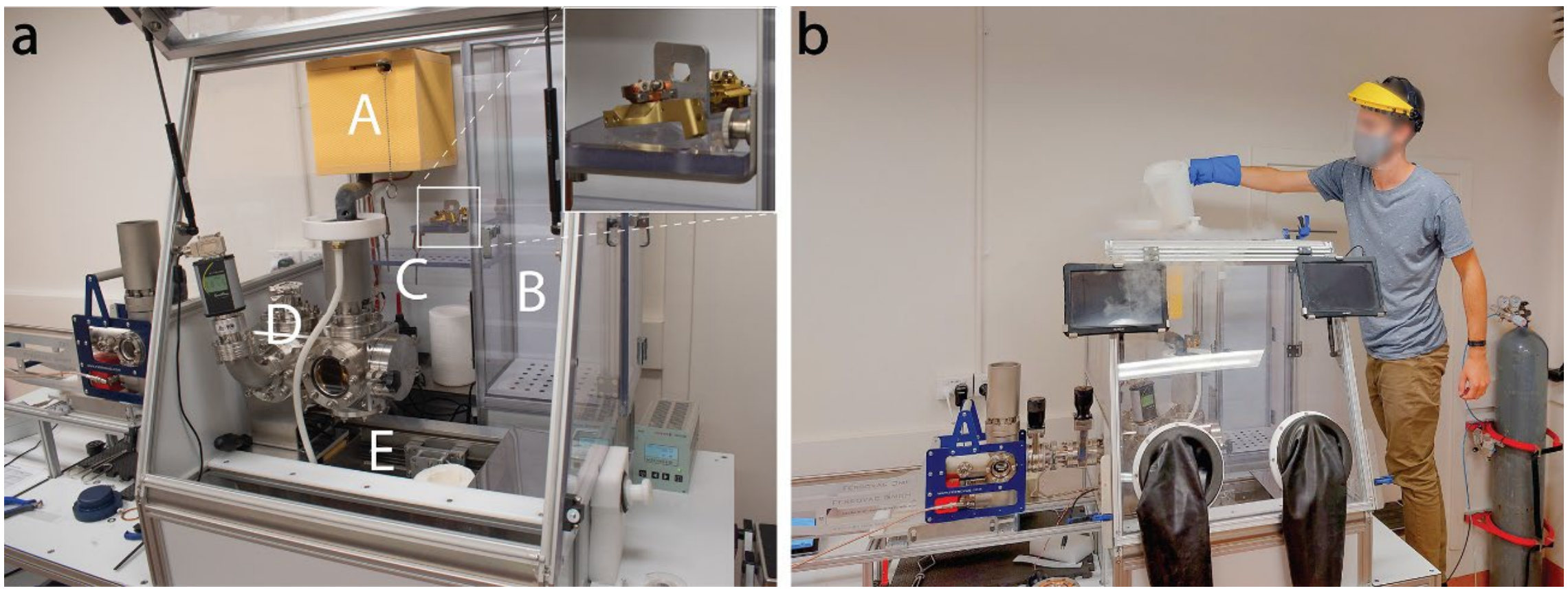

**Figure S12.** (a) Glovebox with hatch open. Inset shows APT samples pre-loaded on cryo-pucks stored on a shelf. (b) Glovebox with hatch closed and operator filling up the top liquid nitrogen dewar.

- Liquid nitrogen supply:

The main liquid nitrogen supply consists of a dewar (A) topped up manually by the operator through the outside of the glovebox, as illustrated in **Figure S12b**. This main dewar is equipped with a flush system to supply the next dewar, located on top of the HVI chamber (D). This intermediate dewar is used to cool down the stage inside the HVI. The white ring on top of the intermediate dewar is an overflow collection ring which redirects the liquid nitrogen down to the liquid nitrogen bath (E) through the white pipe. Other possible liquid nitrogen supply systems also include a pre-filled Styrofoam cylinder introduced inside the glovebox prior to closing the hatch and stored on shelf (C) and a plug on the side of the glovebox (**Figure S12a**, bottom right corner) which, once removed, allows the direct fill of the liquid nitrogen bath from the outside.

- Gaseous nitrogen supply:

Fitted at the bottom of the instrument, to enable and maintain a low humidity and oxygen level to ensure the success of plunge-freezing experiments by preventing the formation of frost.

- Interlock:

The interlock (B) has two doors: one that opens toward the inside of the glovebox, and one than that opens to the outside. It enables the transfer of elements in or out of the glovebox once the hatch has been closed and while the glovebox operator has their hands in the gloves. The top of the interlock is fitted with one direction valves to let the gas inside the glovebox escape.

- Stainless steel high-vacuum interlock chamber:

This high vacuum interlock chamber constitutes the interface between the APT samples prepared/treated inside the glovebox, and the inside of the cryo-vacuum transfer vessel located outside the glovebox, used to transfer APT samples to other instruments at cryogenic temperatures and under vacuum. The specifics of the cryo-vacuum vessel will be detailed further. The HVI is fitted with a turbo-pump able to reach pressures in the order of $10^{-8}$ mbar, and a cryo-stage especially fitted for APT pucks, and cooled down by an external dewar, as mentioned earlier. The turbopump is actioned on the outside of the glovebox, via a switch located on the box visible in **Figure S12a** (bottom right corner). On the inside of the glovebox, APT samples are loaded by opening and closing a door, unlocked by releasing a knob. On the outside of the glovebox, the HVI is fitted with a manually operated gate valve and a docking flange (KF type). The docking flange is the connecting interface with the cryo-vacuum transfer vessel. APT samples are removed from the HVI by opening and then closing the gate valve.

- Liquid nitrogen bath:

The liquid nitrogen bath is at the end of the liquid nitrogen supply chain which starts at the main dewar. Liquid nitrogen is supplied to the bath via the white pipe visible in **Figure S12a**. Once filled with liquid nitrogen, the bath may be used to plunge freeze samples. The bath is also equipped with a heater that enables to boil the liquid nitrogen to generate more gaseous nitrogen.

- Heaters:

Several heaters have been placed on peripheral areas of the glovebox to mitigate cooling from the cold glovebox atmosphere. As mentioned earlier, the liquid nitrogen bath bottom surface is equipped with a heater. The front wall located below the hatch is also equipped with a heater to prevent the operator from being in prolonged contact with cool surfaces while operating the instrument. This heater is slightly protruding inside the glovebox, offering a flat ledge on which items required to be kept warm may be stored (e.g., puck handle, samples, etc)

- Metrics reading:

The glovebox is equipped with an oxygen detector with a ppm sensitivity. The oxygen gauge is located between the HVI and the liquid nitrogen bath. A humidity sensor may be added inside the glovebox before closing the hatch to monitor the humidity level throughout the experiment.

- Additional equipment:

The glovebox is equipped with other additional features to conduct other types of specimen preparation such as hydrogen charging (Chen, et al., 2021).

In order to minimise the risk of frost production during a plunge-freezing operation, it is necessary to maintain humidity as low as possible. This can be achieved by constant monitoring of oxygen and humidity level. As mentioned earlier, an oxygen gauge monitors the oxygen level with ppm sensitivity, however there is no humidity sensor fitted inside the glovebox. A stand-alone humidity sensor device may be introduced in the glovebox to monitor the humidity level throughout an experiment.

The atmosphere inside the glovebox is kept clean of any oxygen and moisture by a constant flow of gaseous nitrogen. This can be achieved by either ensuring the liquid nitrogen inside the glovebox is constantly boiling and therefore evaporating gaseous nitrogen, and/or by supplying gaseous nitrogen directly via the dedicated pipe mentioned earlier. An easy way to produce a lot of gaseous nitrogen is to flush the main liquid nitrogen dewar which will in turn fill the intermediate dewar, then the liquid nitrogen bath. Provided the bath is kept at a warm enough temperature (using a heater), the liquid nitrogen supplied to the bath will keep boiling, and therefore produce gaseous nitrogen, thus maximising the dryness of the atmosphere inside the glovebox.

**Operating the glovebox (plunge-freezing)**

In its idle state the glovebox has its hatch open, oxygen sensor is off, HVI is pumped down, and turbopump is on. Unless in use, the VCT suitcase may remain docked to the glovebox, gate valve closed, and ion pump plugged and running. All the equipment is at room temperature.

At the beginning of an experiment, all the instruments and tools necessary to plunge-freeze the samples need to be introduced inside the glovebox. All items used during our experiments are listed below:

- APT tips mounted on cryo-pucks
- Reverse piggyback puck
- Single puck-slot storage block
- Puck handle (stored on top of the front heater)
- “Prybar” to open the HVI door upon venting without risking tearing the rubber gloves apart due to contact with a cold metal surface

- Tall Styrofoam cylinder filled with liquid nitrogen
- Small Styrofoam cup to carry out plunge-freezing
- Thermometer/humidity sensor device

A piggyback puck is a type of APT puck which can be handled like any normal puck, and on which is mounted a copper mass with a slot for another APT puck, as shown and described in (Stephenson, et al., 2018; Stender, et al., 2022). The reverse piggyback puck is the same, except for the puck slot in the copper mass that is rotated by 180° so that the piggyback puck can be inserted in the HVI from the glovebox side, and the APT puck it carries can be retrieved from the VCT suitcase side (opposite side). Before introducing the cryo-pucks inside the glovebox, the first sample/puck to be plunge-frozen is to be mounted on the reverse piggyback puck, itself mounted on one of the puck storage blocks, while the other sample/puck is to be stored on the remaining puck storage block. This precaution removes the need to handle any of the sample other than during the plunge-freezing process, thus avoiding unnecessary risk of damage. Once all the tools are introduced, the humidity sensor may be switched on, and the hatch may be locked shut.

The glovebox requires two people to operate safely. After closing the hatch, operator #1 may put the rubber gloves on while operator #2 pours the liquid nitrogen into the main dewar to (1) start cooling down the cryo-stage inside the HVI and (2) start producing nitrogen gas inside the chamber to reduce the humidity and oxygen level. Once the main dewar reaches near full capacity, it will overflow with liquid nitrogen and start filling the intermediary dewar, as illustrated in **Figure S13**. To better control the liquid nitrogen flow between dewars, it is preferable to slowly pour into the main dewar and release to the main dewar by flushing rather than overflowing. These operations are repeated until the nitrogen in the intermediary dewar stops boiling, after which the latter is overflown to start pouring nitrogen towards the liquid nitrogen bath. These actions are then repeated until the humidity and oxygen level are suitable for cryotransfer, typically 0.0 % and below 2000 ppm, respectively. It takes approximately one hour from the first pouring of liquid nitrogen down the main dewar to reach suitable humidity and oxygen level.

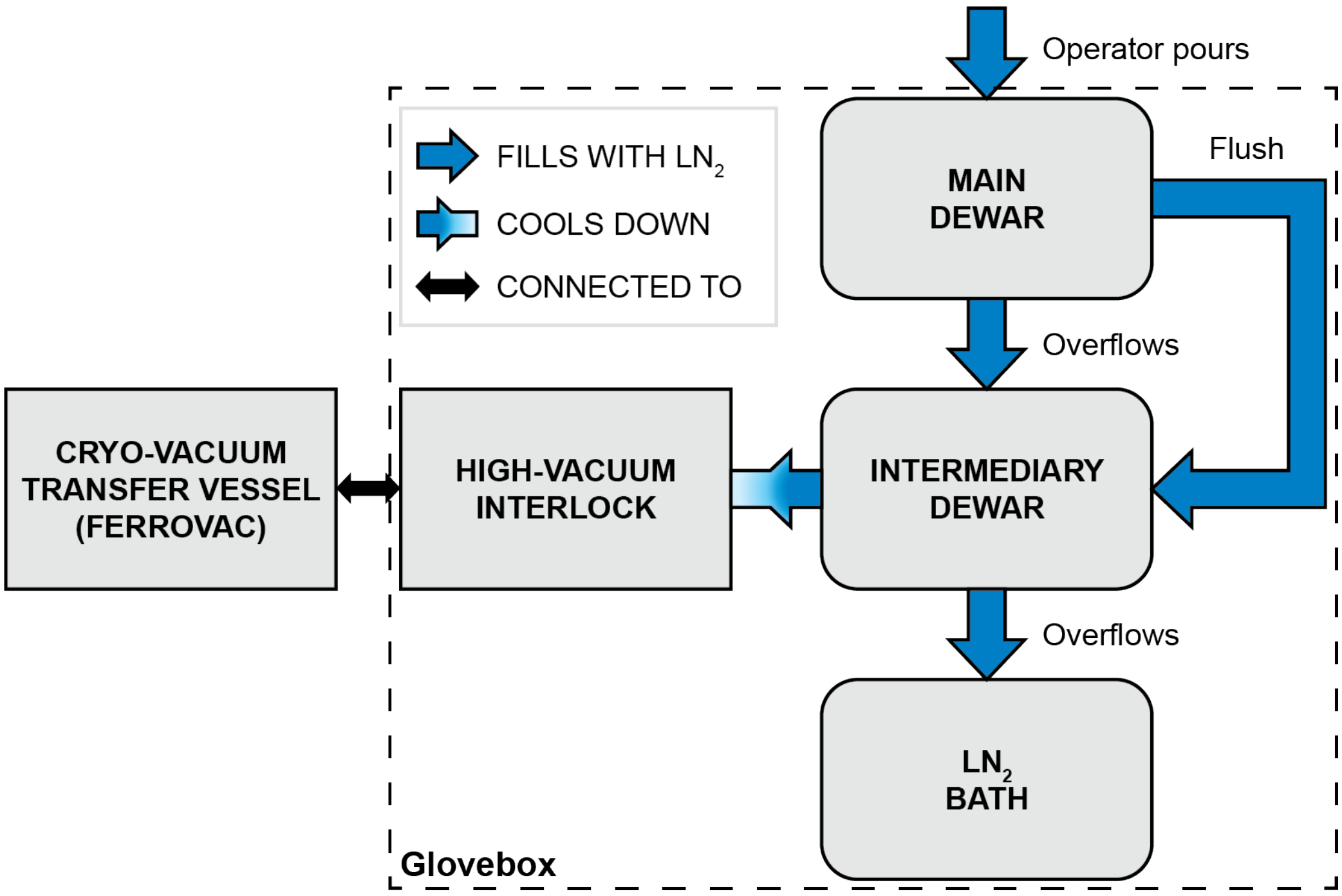

**Figure S13.** Schematic of the liquid nitrogen flow inside the glovebox

Once the thresholds are reached and provided the cryo-transfer vessel is ready (see Vacuum cryogenic transfer suitcase section), the first step of the cryo-transfer protocol may be started – i.e., plunge-freezing, and subsequent storing into HVI.

First, the HVI is vented by operator #2. While the HVI is venting, the tall Styrofoam cylinder is used to pour the remaining liquid nitrogen into the small Styrofoam cup, in which the plunge freezing process will be conducted. The liquid nitrogen will start to boil as soon as it is poured into the Styrofoam cup, therefore the latter needs to be topped up again once boiling has stopped. The following steps need to be carried out as quick as possible in order to achieve a reasonable balance between avoiding humidity to move to the HVI and avoiding damaging the samples by physical damage or frost build up.

Once the HVI is completely vented, operator #1 uses the "prybar" to open the door, connects the puck handle to the reverse piggyback puck and then plunge-freezes the sample until the nitrogen stops boiling. The operator then removes the puck from the liquid nitrogen and quickly insert it on the cryo-stage inside the HVI, and finally shuts the door, following that operator #2 switches the turbopump on. The glovebox may then be opened if there are no samples remaining to be transferred.

In case there is another sample, the glovebox is to remain shut, and the liquid nitrogen level/flow needs to be monitored to keep low humidity and oxygen level until the next plunge-freezing/transfer operation. The process for any subsequent sample is identical although everything inside the glovebox is already pre-cooled. This comes as an advantage because much time is spared due to not having to cool all cryogenic parts to liquid nitrogen temperature. However, it also brings new challenges, especially regarding the non-cryogenic parts (i.e., handling tools, samples) because if their temperature drops too low (namely below 0 °C) they might be subject to frost build up due to the uncontrollable (albeit small) amount of humidity remaining in the chamber. To avoid this, the heaters are to be used, relocating critical items to be in their vicinity, so that these items do not reach critically low temperatures.

After all samples have been transferred (1) to the HVI, and then (2) out of the HVI into the cryo-vacuum transfer vessel, the glovebox may be opened and cleaned if required. The heater located near the main dewar may be activated to accelerate the evaporation of liquid nitrogen and subsequent warming up of all other parts inside the glovebox.

## Cryo-FIB

The cryo-FIB is a Zeiss Auriga FIB-SEM equipped with a custom cryo-transfer interface and cryo-stage (Microscopy Solutions) that has an operational cryo-temperature of ~ -126 °C (~147 K). The instrument is shown in **Figure S14a** in the centre, the cryo-temperature monitor is on the left, and the operating computer is on the right. The cryo-transfer interface is made of a cryogenically cooled high-vacuum loadlock connected to the main instrument chamber by a gate valve. The instrument is cryogenically cooled by two liquid nitrogen dewars located near the main chamber (arrow in **Figure S14a**) and near the loadlock (top left arrow **Figure S14c**), respectively. The loadlock is also equipped with a dock compatible with a Ferrovac VCT suitcase. The instrument is represented in **Figure S14** with (a) and without (b) the VCT suitcase docked.

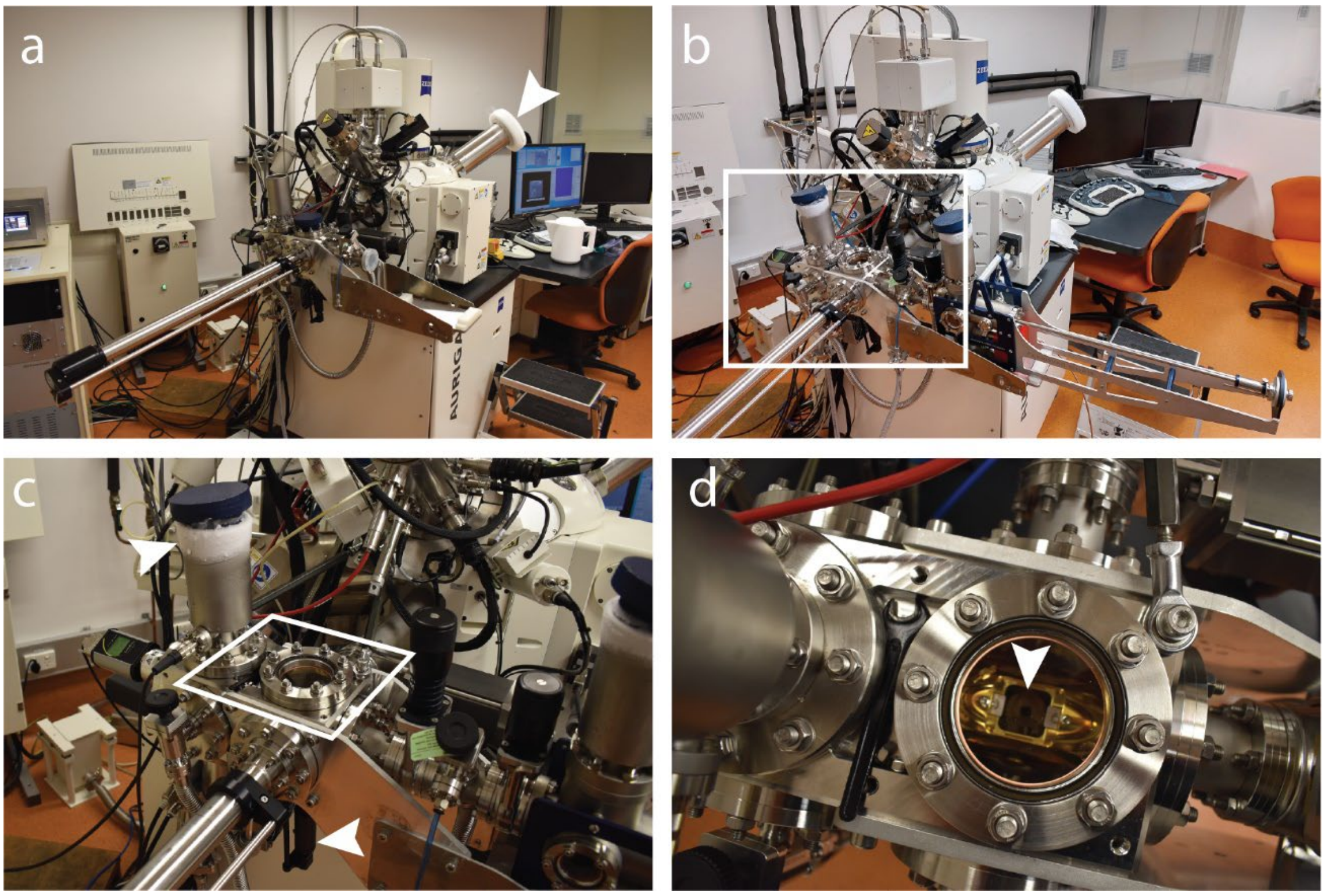

**Figure S14.** Cryo-FIB instrument (a) Cryo-transfer interface. (b) Cryo-transfer interface with VCT suitcase docked. (c) Close-up of the cryo-transfer interface. Top left arrow shows the loadlock dewar. Bottom arrow indicates the cryo-puck loading stage handle. (d) Close-up on the load lock. Arrow shows the cryo-puck loading stage.

**Cryo-transfer using the VCT suitcase**

Prior to docking the VCT containing an APT sample, every cryo-component needs to be cooled to cryogenic temperature and stable. Due to the high thermal mass of the system, it is preferable to pre-cool the instrument at least one day before the experiment by regularly topping up the liquid nitrogen level in the dewars. Readings of the cryo-stage temperature were collected throughout the experiment and are summarised in **Figure S15**.

When the loadlock is at ~ -186 °C (~ 87 K) and the cryo-stage is at ~ -126 °C (~147 K) the VCT suitcase may be docked to the FIB-SEM using either a CF or KF flange. Then, the space between the loadlock and the VCT suitcase, in between the two gate valves visible in **Figure S14c,** needs to be pumped down. This is achieved with a two-step process where a vacuum pump first brings the pressure down to ~ $10^{-2}$ mbar. Then a hose connected to the loadlock is gradually opened to reach high vacuum using the turbopump. This step is one of the most critical of the cryo-transfer process because of the risk of overloading the turbo-pump if the valve is opened too fast. Once equilibrium is reached between the loadlock and the space near the flange, the gate valves on both sides may be opened and the cryo-transfer may begin.

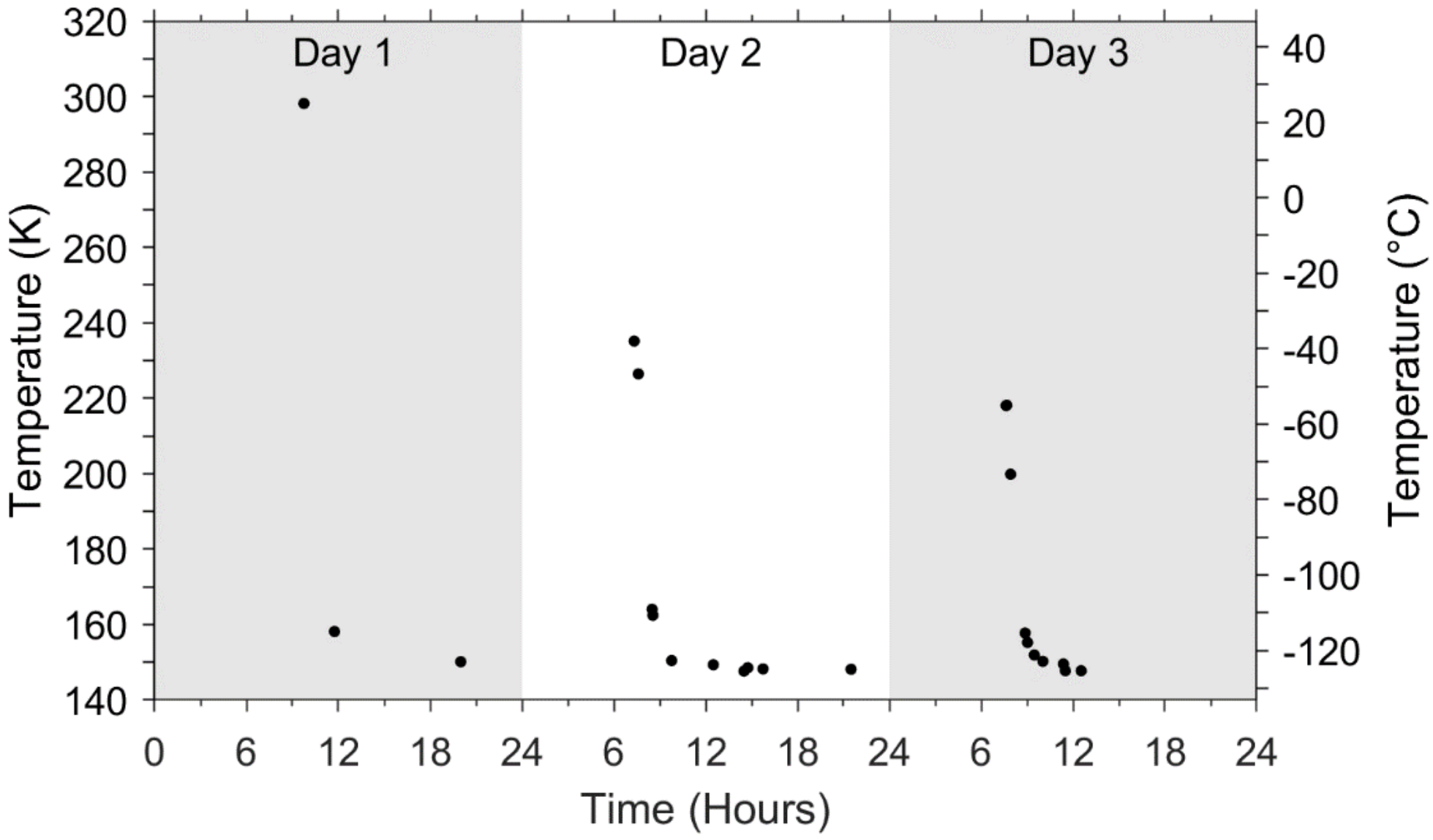

**Figure S15.** Cryo-stage temperature evolution

The central piece of the transfer system is a rotating stage, indicated by the arrow in **Figure S14d**, where the cryo-shield can also be seen. The stage can be moved up and down, and rotated using the handle shown in **Figure S14c**. To transfer an APT sample puck from the VCT to the main chamber of the SEM, the puck must be transferred to the rotating stage using the handling rod of the VCT suitcase. Then, the rod is retracted, and the VCT gate valve closed. The stage is then rotated and positioned so that the horizontal transfer rod of the instrument can be used to retrieve the puck and insert it on the cryo-stage inside the main chamber. Once the sample is transferred, the horizontal rod is retrieved, and all the gate valves are closed. The entire transfer process must be conducted as fast as possible, and all steps must be carried out in a precise order to minimise heat transfer at any stage of the process.

**Cryo-FIB milling**

The purpose of cryo-annular milling was to remove the graphene layer to study and graphene-free sample in the atom probe. However, some frost had appeared during the entire process, as shown in **Figure S16**. Although it is impossible to determine precisely at which stage this contamination occurred, it is most likely after removing the sample from liquid nitrogen and before it was pumped in the HVI. Cryo-annular milling was used to remove the frost at the tip apex of the specimen along with the graphene. It should be noted that the imaging resolution was drastically reduced because of vibrations caused by the turbopump, which could not be turned off during the milling process in order to maintain optimal conditions inside the load lock. This reduced image quality made it more challenging to analyse the surface at the apex of the APT needle, and therefore to accurately target the graphene layer using visual cues. Nonetheless, annular milling was carried out by successive steps with decreasing milling current and milling pattern radii. Once the frost and graphene layer were removed, the cryo-transfer process described in the previous section was reversed to transfer the APT sample back to the VCT suitcase.

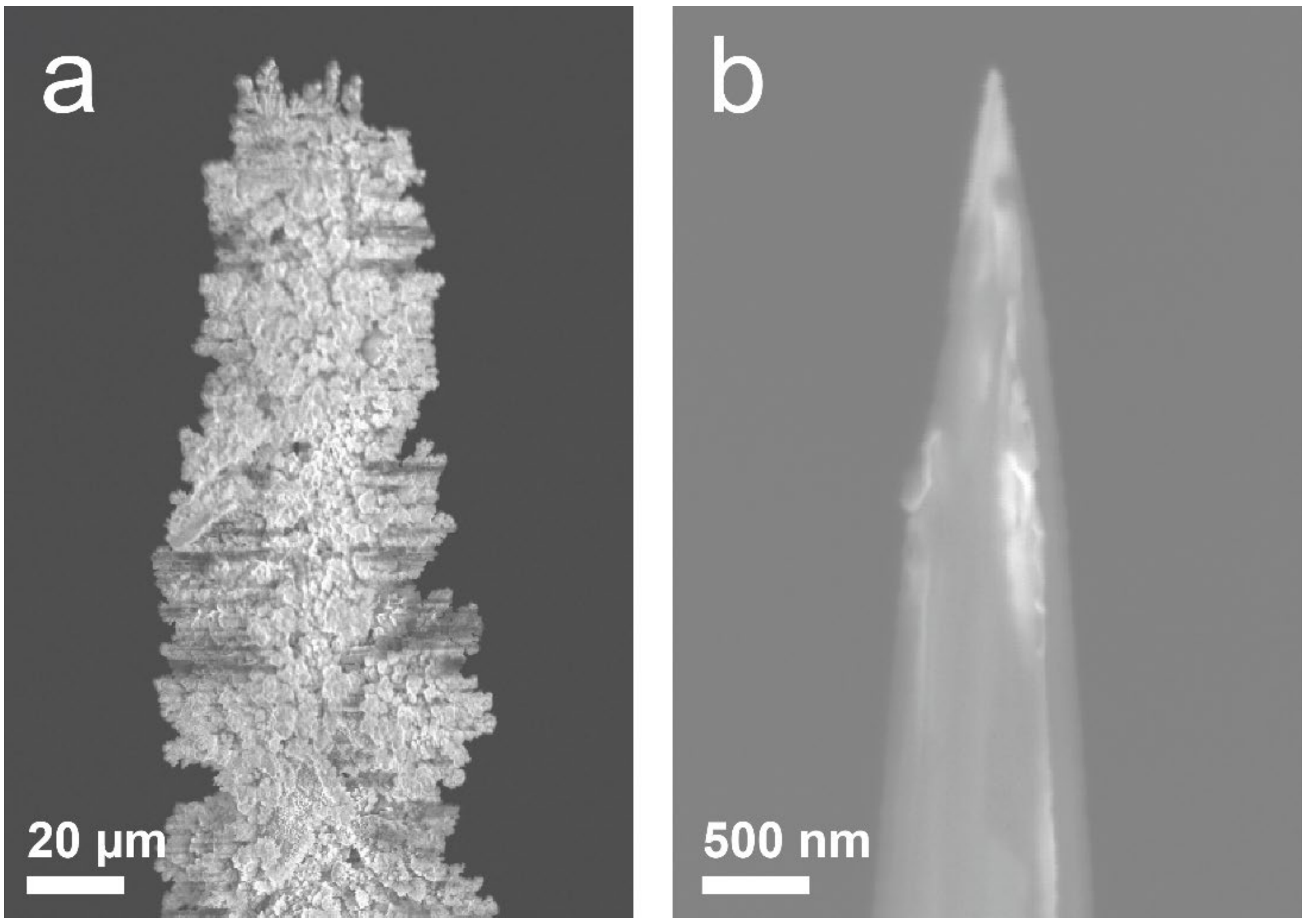

**Figure S16.** SEM images of a cryo-prepared APT sample (a) before annular milling, and (b) after annular milling.

## Cryo-APT

### Equipment

The atom probe instrument used in this workflow was a LEAP 4000 X Si (Cameca Instruments) equipped with the following cryo-accessories:

- Special load lock with a docking interface compatible with the Ferrovac suitcase
- Cryo-carousel with a polyetheretherketone (PEEK) slot suitable with cryogenically cooled APT pucks
- Piggyback puck designed to hold another APT puck
- Cryo-pucks fitted with a PEEK extremity to prevent any room-temperature handling element from heating up its cryogenically cooled copper mass

**Cryo-transfer and acquisition**

Prior to connecting the VCT suitcase to the atom probe, the piggyback puck must be inserted inside the analysis chamber of the instrument to be cooled down to 50 K, which takes approximately 30 min. Additionally, at least one of the carousels inside the instrument must be fitted with a PEEK slot.

The process of docking the VCT suitcase to the LEAP is very similar to that described in the Cryo-FIB section. Upon docking, the space between the instrument gate valve and the VCT gate valve interface needs to be pumped down, which is achieved by slowly bleeding gas into the diaphragm pump that backs the loadlock turbopump. The valve needs to be opened extremely slowly to avoid gas going backwards through the turbo pump, which could crash the pump and have dramatic consequences for the instrument and the rest of the experiment. After equilibrium has been reached, the loadlock gate valve may be opened, and the transfer process may be started.

Although the transfer process can be achieved by one single operator, is it more efficiently carried out by two persons. Additionally, the transfer process must be performed as fast as possible to minimise sample heating. The first step is to retrieve the piggyback puck from the analysis chamber onto the PEEK slot of the carousel, which is then raised up to the loadlock. The carousel needs to be rotated so that the piggyback is facing the VCT suitcase gate valve. After opening the gate valve, the sample may be inserted onto the piggyback puck from the VCT suitcase. The upper carousel may then be lowered to the buffer chamber while the second operator closes the VCT suitcase gate valve. The sample is then removed from the piggyback puck using the LEAP horizontal transfer rod and inserted into the analysis chamber, following which the APT acquisition may be carried out in the same way as a conventional sample. It should be noted that at this stage, the piggyback puck is still at cryogenic temperature while the rest of the buffer chamber is at room temperature, which causes the piggyback puck to act as a cold trap. To avoid the frost built up at its surface from sublimating and contaminating the vacuum, it must be removed from the buffer chamber, and if possible, from the instrument altogether.

From plunge-freezing to APT acquisition, there are many steps during which the samples can be damaged or contaminated. Between temperature and vacuum management, cross-platform sample handling, and the sometimes-time-consuming pre-experiment instrument cooling, the whole process must be conducted with great care. In this series of experiments, all cryo-transferred samples had frost build up at some stage. Although it is suspected that this happened during the very first step (plunge-freezing), it is hard to precisely determine the real cause. Whilst the frost may spare the apex of the tip, as was the case for one of the samples, it may also damage the sample and affect the APT experiment, as shown in **Figure 2** where a chunk of frost can be seen enveloping the specimen tip.

# Section 4: Successive cryo-workflows steps

## Day 1: Electropolishing

- The samples were electropolished at Deakin University
- Number of samples prepared: 16
- Samples type: tungsten wire crimped in copper tube.
- Method: Fine electropolishing with 5% NaOH solution using 2 V AC current.
- Samples were labelled as follows: 220314_LT+FE_WE-EP-Syd*** 001 to 016

## Day 2: SEM imaging

- Samples were imaged using FEI Quanta 3D FEG SEM (Deakin University).
- All 16 samples were imaged at 5 kV, 0.42 nA, 45 deg tilt, with decreasing magnifications of 100k, 50k, 20k.
- The diameter was measured for each sample.
- **Yield: 15/16**

## Days 3 to 4: Pre-run + graphene coating

### Pre-run

| Sample name | Pre-run | | | | Graphene-coating | SEM imaging |
|---|---|---|---|---|---|---|
| | R04_234** | kV | Ion count | Outcome | | |
| Syd001 | 23472 | 5.9 | 220880 | Good | 5x | Good |
| Syd002 | 23486 | 4 | 181941 | Good | 7x | Bad |
| Syd003 | 23485 | 2 | 172473 | Good? | 4x | Bad |
| Syd004 | 23473 | 4.7 | 144918 | Good | 7x | OK |
| Syd005 | 23474 | 5 | 942016 | Average | 7x | OK |
| Syd006 | 23484 | 3.5 | 123681 | Good | 6x | OK |
| Syd007 | 23475 | 6.5 | 706254 | Fractured | | |
| Syd008 | Bent | | | | | |
| Syd009 | 23480 | 4.9 | 124103 | Good | 6x | Good |
| Syd010 | 23482 | 6.3 | 254916 | Good | 7x | Good |
| Syd011 | 23487 | 6.7 | 18544 | Blunt | | |
| Syd012 | 23483 | 2.2 | 198078 | Great | 8x | Good |
| Syd013 | 23488/89 | 6.5 | 140196 | Blunt | | |
| Syd014 | 23477 | 8.5 | 127597 | Blunt | | |
| Syd015 | 23478 | 5.6 | 132224 | Good | 6x | OK |
| Syd016 | 23481 | 3.7 | 247017 | Good | 5x | Bad |

- **Pre-run yield: 11/15**
- **Graphene-coating yield: 8/11**

## Days 4 to 5: SEM imaging

Instrument: Zeiss Sigma.

Parameters:

| SEM IMAGING PARAMETERS | |
|---|---|
| Accelerating voltage | 5 kV |
| Working distance | ~ 5 mm |
| Imaging mode | SE2, InLens |
| Magnifications | x10k, x100k, x200k + relevant magnificationss based on specimen geometry |
| First session (15/03) | Samples Syd001, Syd004, Syd005, Syd009, Syd015, Syd016 |
| First session (17/03) | Samples Syd002, Syd003, Syd006, Syd010, Syd012 |
| **ASSESSMENT OF GRAPHENE COATING** | |
| Good samples | Syd001, Syd009, Syd010, Syd012 |
| OK samples | Syd004, Syd005, Syd006, Syd015 |
| Failed samples | Syd002, Syd003, Syd016 |

## Day 5 to 7: Cryo-workflows

### Summary of workflows

**Scenario A steps:**

| Step | Instrument |
|---|---|
| Plunge-freezing | Glovebox |
| Cryo-transfer to LEAP 4000 XSi | Ferrovac suitcase |
| APT analysis | LEAP 4000 XSi |

**Scenario B steps:**

| Step | Instrument |
|---|---|
| Plunge-freezing | Glovebox |
| Cryo-transfer to Cryo-FIB | Ferrovac suitcase |
| SEM imaging + FIB milling of graphene cap | Zeiss Auriga FIB-SEM |
| Cryo-transfer to LEAP 4000 XSi | Ferrovac suitcase |
| APT analysis | LEAP 4000 XSi |

**APT samples**

<table>
<tr><td>Sample 1</td><td colspan="2">220314_LT+FE_W-Syd004_GC</td></tr>
<tr><td rowspan="16">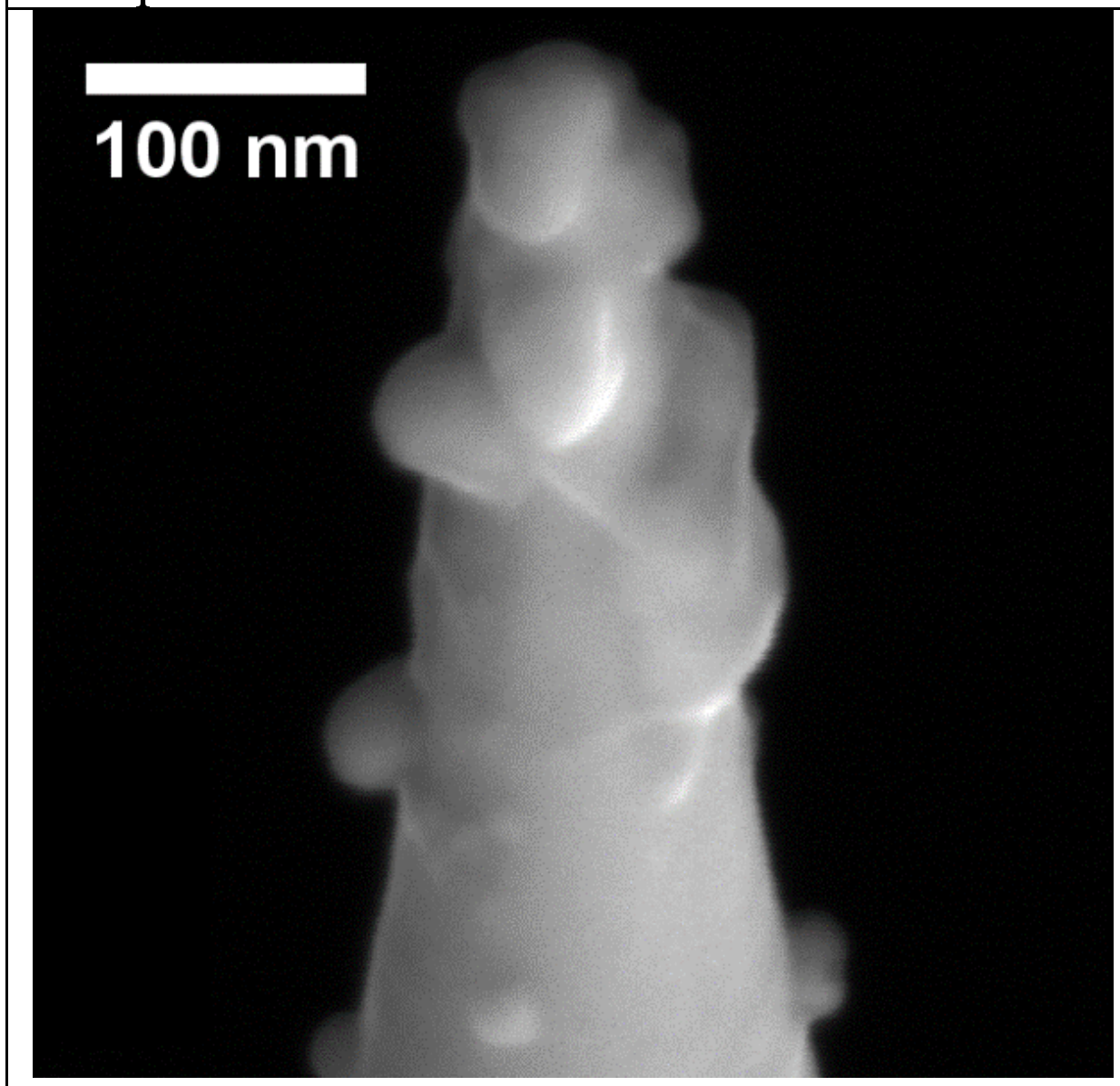
</td><td>Cryo-transfer date</td><td>16/03/2022</td></tr>
<tr><td>Acquisition date</td><td>16/03/2022</td></tr>
<tr><td>RHIT file</td><td>R18_61143</td></tr>
<tr><td>Pre-run voltage</td><td>4.7 kV</td></tr>
<tr><td>Puck/location</td><td>C020</td></tr>
<tr><td>Needle length</td><td>12 mm</td></tr>
<tr><td>Acquisition mode</td><td>Laser</td></tr>
<tr><td>Temperature</td><td>50 K</td></tr>
<tr><td>Pulse energy</td><td>20 pJ</td></tr>
<tr><td>Detection rate</td><td>0.1 %</td></tr>
<tr><td>Pulse rate</td><td>125 kHz</td></tr>
<tr><td>Vacuum</td><td>2.23E-11</td></tr>
<tr><td>Starting voltage</td><td>500</td></tr>
<tr><td>Voltage control</td><td>Manual</td></tr>
<tr><td>Total ions</td><td>3.83E5</td></tr>
<tr><td>Final voltage</td><td>4.3 kV</td></tr>
</table>

<table>
<tr><td>Sample 2</td><td colspan="2">220314_LT+FE_W-Syd009_GC</td></tr>
<tr><td rowspan="16">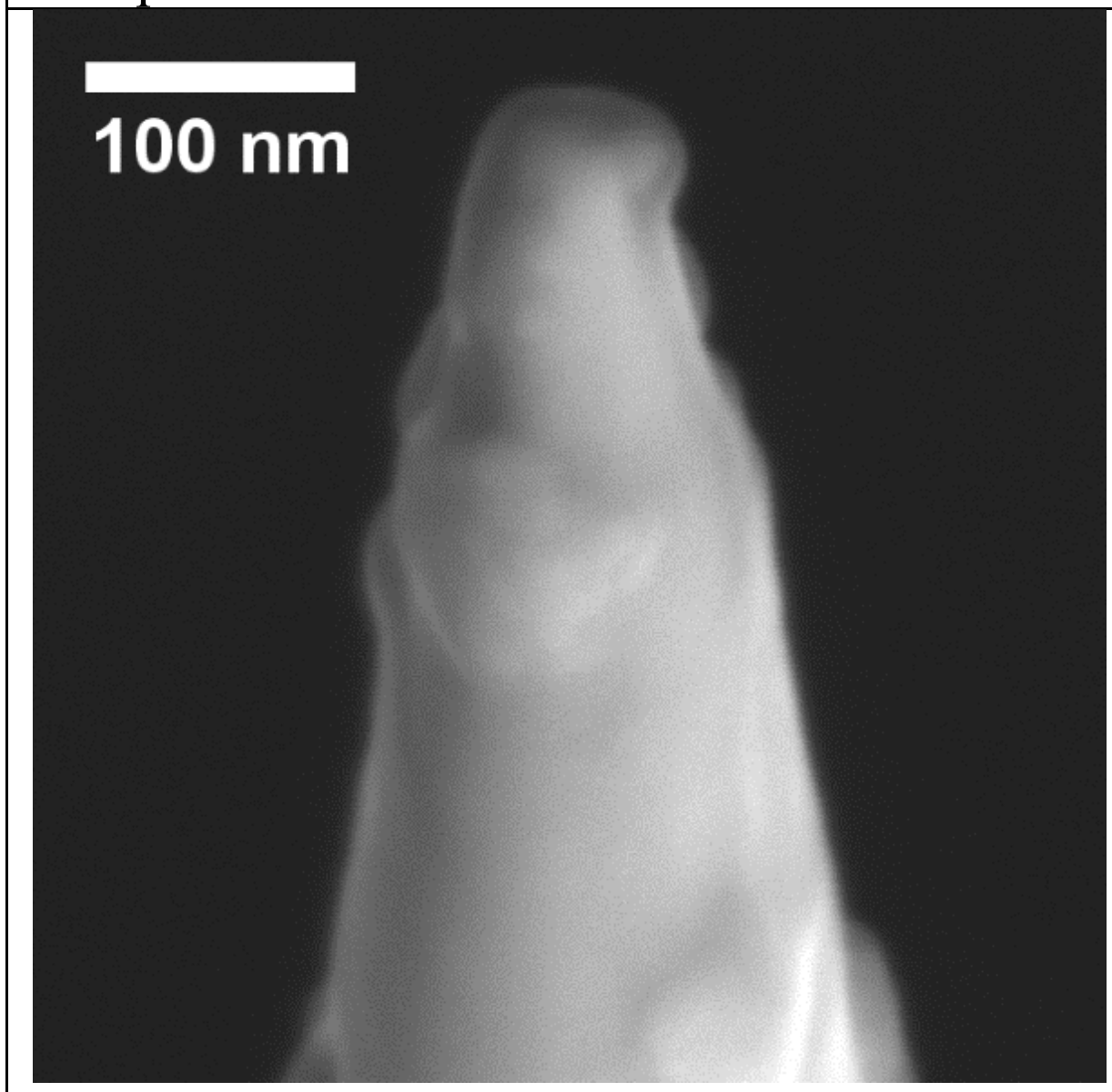
</td><td>Cryo-transfer date</td><td>16/03/2022</td></tr>
<tr><td>Acquisition date</td><td>17/03/2022</td></tr>
<tr><td>RHIT file</td><td>R18_61147</td></tr>
<tr><td>Pre-run voltage</td><td>4.9 kV</td></tr>
<tr><td>Puck/location</td><td>C025</td></tr>
<tr><td>Needle length</td><td>14 mm</td></tr>
<tr><td>Acquisition mode</td><td>Voltage</td></tr>
<tr><td>Temperature</td><td>50 K</td></tr>
<tr><td>Pulse fraction</td><td>20 %</td></tr>
<tr><td>Detection rate</td><td>0.1 %</td></tr>
<tr><td>Pulse rate</td><td>125 kHz</td></tr>
<tr><td>Vacuum</td><td>1.9E-11</td></tr>
<tr><td>Starting voltage</td><td>500</td></tr>
<tr><td>Voltage control</td><td>Manual</td></tr>
<tr><td>Total ions</td><td>3.51E5</td></tr>
<tr><td>Final voltage</td><td>7.2 kV</td></tr>
</table>

| Sample 3 | 220314_LT+FE_W-Syd015_GC | |
|---|---|---|
| 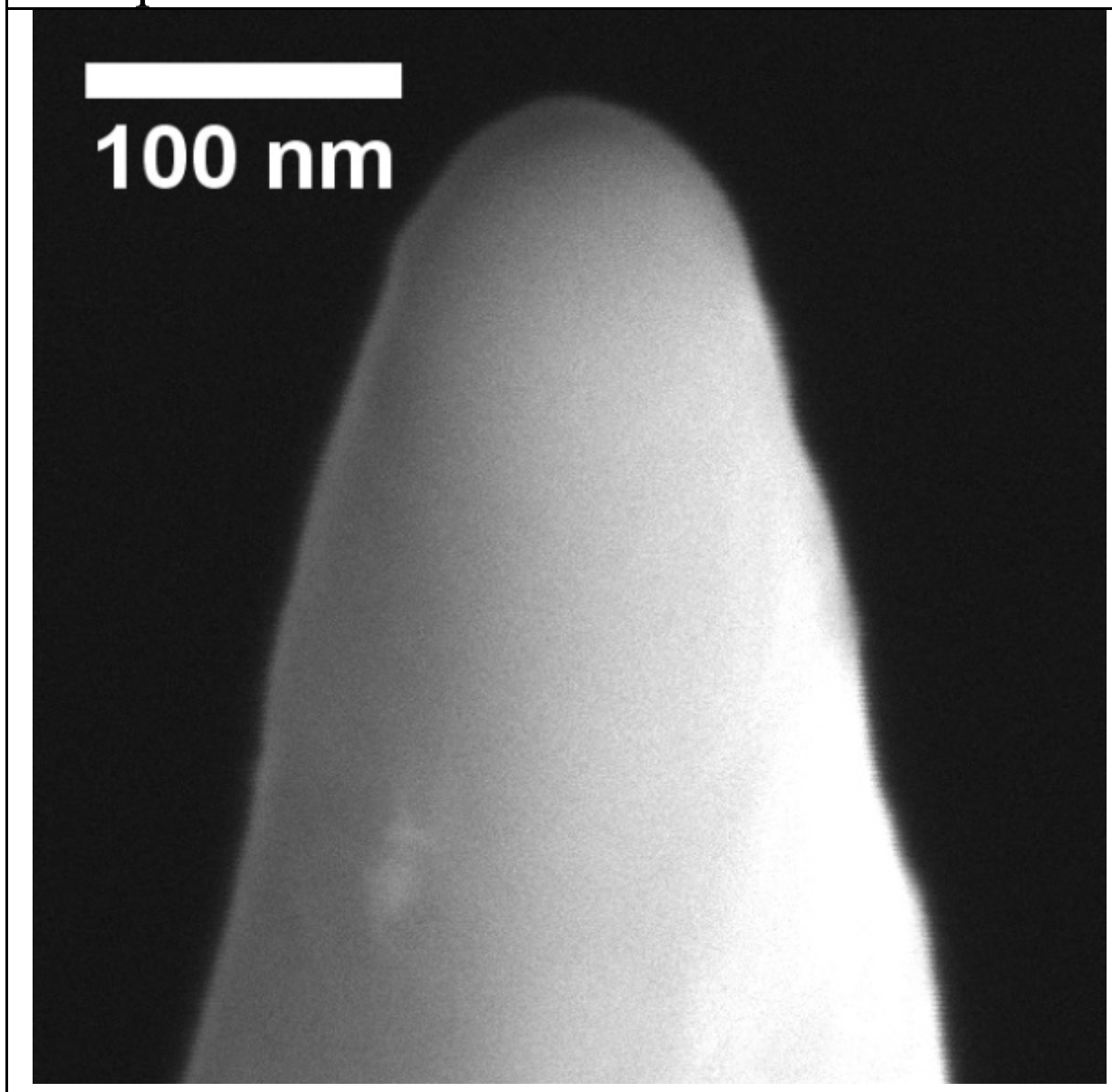  | Cryo-transfer date | 17/03/2022 |
| | Date analysis | 17/03/2022 |
| | RHIT file | R18_61148 |
| | Pre-run voltage | 5.6 kV |
| | Puck/location | C020 |
| | Needle length | 13 mm |
| | Acquisition mode | Laser |
| | Temperature | 50 K |
| | Pulse energy | 20 pJ |
| | Detection rate | 0.1 % |
| | Pulse rate | 125 kHz |
| | Vacuum | 2.2E-11 |
| | Starting voltage | 500 |
| | Voltage control | Manual |
| | Total ions | |
| | Final voltage | |

| Sample 4 | 220314_LT+FE_W-Syd001_GC | |
|---|---|---|
| 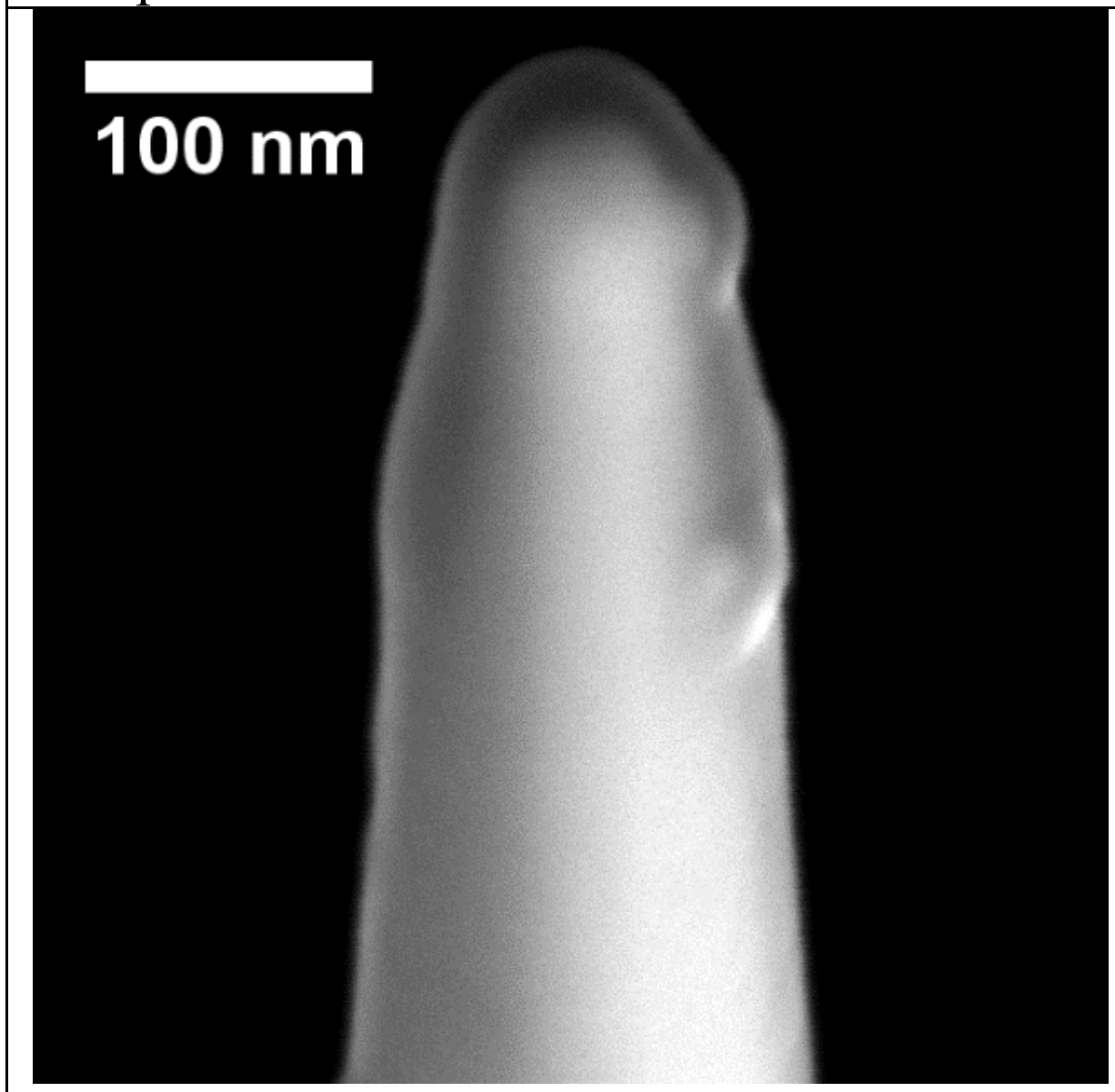  | Cryo-transfer date | 18/03/2022 |
| | Date analysis | 18/03/2022 |
| | RHIT file | R18_61149 |
| | Pre-run voltage | 5.9 kV |
| | Puck/location | C025 |
| | Needle length | 13 mm |
| | Acquisition mode | Laser |
| | Temperature | 50 K |
| | Pulse energy | 20 pJ |
| | Detection rate | 0.1 % |
| | Pulse rate | 125 kHz |
| | Vacuum | 1.9E-11 |
| | Starting voltage | 500 |
| | Voltage control | Manual |
| | Total ions | 4.45E5 |
| | Final voltage | 5.2 kV |

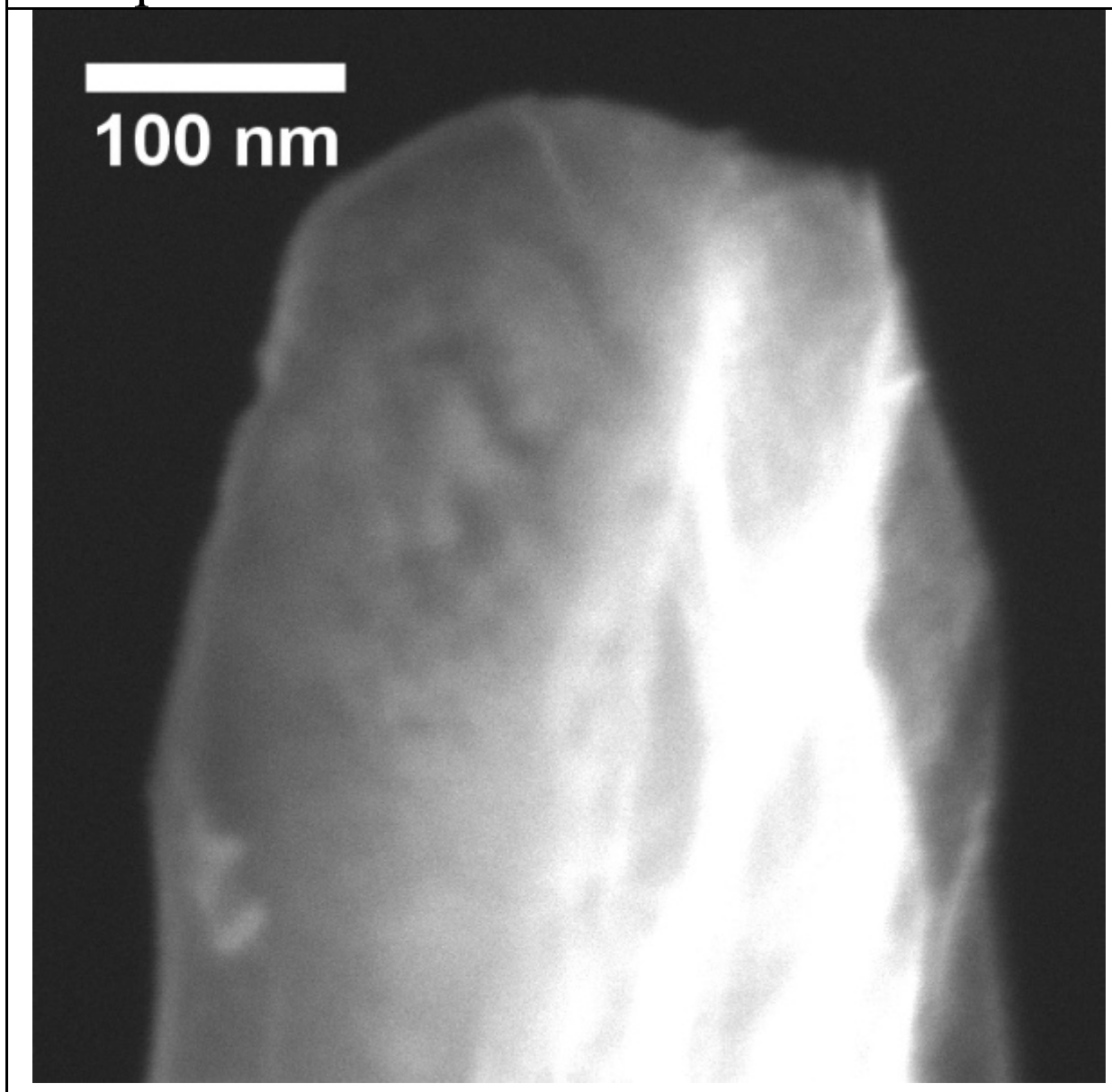

| Sample 5 | 220314_LT+FE_W-Syd012_GC |
| --- | --- |
| Cryo-transfer date | 18/03/2022 |
| Date analysis | 18/03/2022 |
| RHIT file | R18_61151 |
| Pre-run voltage | 2.2 kV |
| Puck/location | C020 |
| Needle length | 14 mm |
| Acquisition mode | Voltage |
| Temperature | 50 K |
| Pulse fraction | 20 % |
| Detection rate | 0.1 % |
| Pulse rate | 125 kHz |
| Vacuum | |
| Starting voltage | 500 |
| Voltage control | Manual |
| Total ions | 1.07E5 |
| Final voltage | 3.3 kV |

## Day 8: Room temperature specimen transfer

Three specimens were left in Sydney to be analysed following regular, room-temperature transfer to the atom probe: Syd005, Syd 006 and Syd007.

| Sample name | Location | Pre-run | | | | Graphene-coating | APT run |
| --- | --- | --- | --- | --- | --- | --- | --- |
| | | R04_234** | kV | Ion ct | Outcome | | |
| Syd005 | S01 PE | 23474 | 5 | 942 | Average | 7x | Bent |
| Syd006 | S01 PF | 23484 | 3.5 | 123k | Good | 6x | Voltage |
| Syd010 | S01 PJ | 23482 | 6.3 | 250k | Good | 7x | Laser |

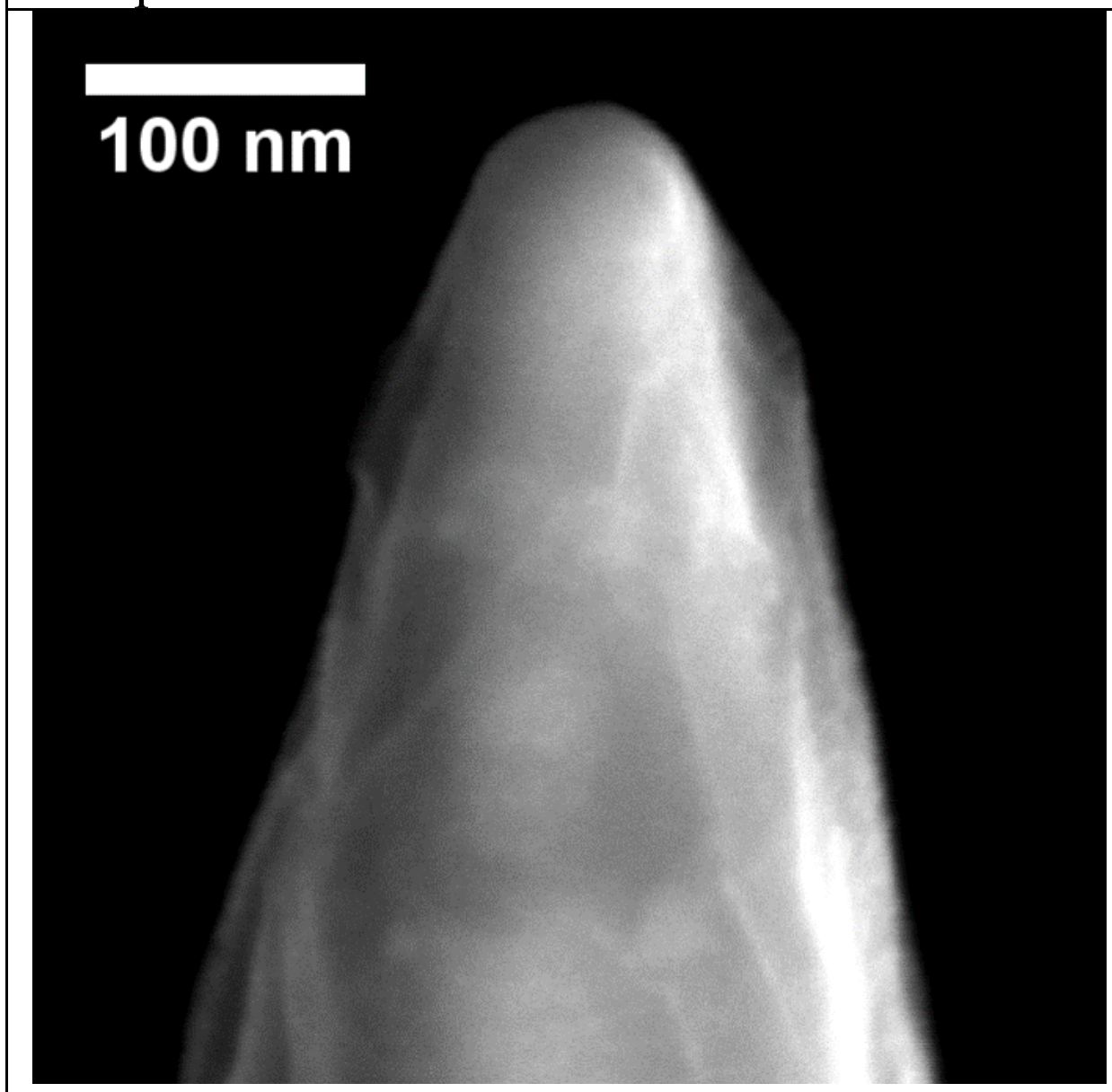

| Sample 1 | 220314_LT+FE_W-Syd010_GC |
| --- | --- |
| RT transfer date | 30/03/2022 |
| Date analysis | 30/03/2022 |
| RHIT file | R18_ |
| Pre-run voltage | 6.3 kV |
| Puck/location | |
| Needle length | |
| Acquisition mode | Laser |
| Temperature | 50 K |
| Pulse energy | 20 pJ |
| Detection rate | 0.1 % |
| Pulse rate | 125 kHz |
| Vacuum | |
| Starting voltage | 500 |
| Voltage control | Manual |
| Total ions | 5E5 |
| Final voltage | 7.8 kV |

| Sample 2 | 220314_LT+FE_W-Syd005_GC | |
|---|---|---|
| 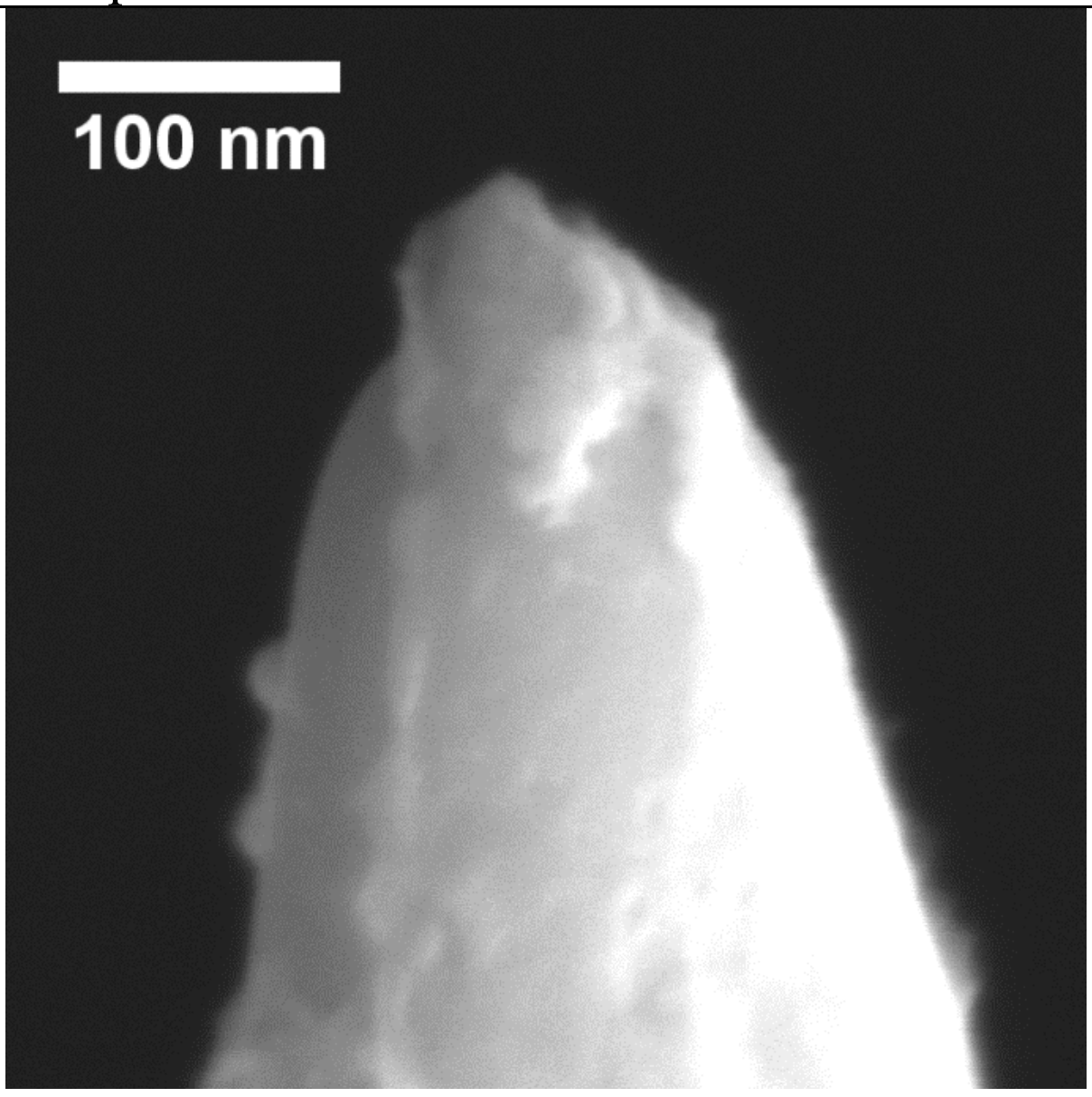 | RT transfer date | 30/03/2022 |
| | Date analysis | 30/03/2022 |
| | RHIT file | R18_ |
| | Pre-run voltage | 3.5 kV |
| | Puck/location | |
| | Needle length | |
| | Acquisition mode | Voltage |
| | Temperature | 50 K |
| | Pulse fraction | 20 % |
| | Detection rate | 0.1 % |
| | Pulse rate | 125 kHz |
| | Vacuum | |
| | Starting voltage | 500 |
| | Voltage control | Manual |
| | Total ions | |
| | Final voltage | |